# Chiral Epitaxy: Enantioselective Growth of Chiral Nanowires on Low-Symmetry Two-Dimensional Materials

Noya Ruth Itzhak[1]†, Kate Reidy[2]†, Maya Levy-Greenberg[1], Paul Anthony Miller[3], Chen Wei[4], Juan Gomez Quispe[5], Raphael Tromer[6], Olle Hellman[1], Shahar Joselevich[7], Aliza Ashman[1], Lothar Houben[8], Ifat Kaplan-Ashiri[8], Xiao-Meng Sui[8], Olga Brontvein[8], Katya Rechav[8], Laurent Travers[4], Pedro A. S. Autreto[5], Douglas S. Galvão[6], Federico Panciera[4], Oded Hod[9], Leeor Kronik[1], Frances M. Ross[3*] and Ernesto Joselevich[1*]

[1]Department of Molecular Chemistry and Materials Science, Weizmann Institute of Science; Rehovot, 7610001, Israel.
[2] Miller Institute for Basic Research in Science, University of California; Berkeley, California, 94720, USA.
[3]Department of Materials Science and Engineering, Massachusetts Institute of Technology; Cambridge, 02139, USA.
[4]Université Paris-Saclay, CNRS, Centre de Nanosciences et de Nanotechnologies; Palaiseau, 91120, France.
[5]Center for Natural and Human Sciences, Federal University of ABC; Santo André, São Paulo, 09210-170, Brazil.
[6]Applied Physics Department and Center for Computing in Engineering and Sciences, State University of Campinas; Campinas, 13083-959, São Paulo, Brazil.
[7]De Shalit High School; Rehovot, 7624720, Israel.
[8]The Department of Chemical Research Support, Weizmann Institute of Science; Rehovot, 7610001, Israel.
[9]Department of Physical Chemistry, School of Chemistry, The Raymond and Beverly Sackler Faculty of Exact Sciences and The Sackler Center for Computational Molecular and Materials Science, Tel Aviv University; Tel Aviv, 6997801, Israel

*Corresponding author. Email: fmross@mit.edu (F.M.R.);
ernesto.joselevich@weizmann.ac.il (E.J.)
†These authors contributed equally to this work

Chiral crystals exhibit useful handedness-dependent properties, including spin selectivity and circularly polarized light sensitivity, yet controlling which enantiomer forms during synthesis remains a central challenge. Existing approaches utilize molecules in solution to template crystal growth, which restricts processing conditions and introduces organic contaminants incompatible with device fabrication. Enantioselective growth of a chiral crystal on a chiral surface via vapor-phase synthesis ('chiral epitaxy') has not yet been demonstrated. Here, we show chiral epitaxy of aligned tellurium nanowires on a low-symmetry two-dimensional material, ReSe$_2$. *In situ* electron microscopies suggest a mechanism where handedness is determined at nucleation by the interface energy difference between Te enantiomers and the chiral substrate surface. Chiral epitaxy provides a solvent-free, vapor-solid route to homochiral crystals compatible with semiconductor and quantum manufacturing processes.

## Introduction

Since Pasteur's 1848 discovery that certain molecules and crystals, named 'chiral', can have two distinct mirror-image forms, one left-handed and one right-handed, (1) achieving deliberate control over handedness has remained a major challenge, one whose resolution would advance not only our fundamental understanding of handedness-selective ('enantioselective') growth but also the long-term development of chiral devices. Recently, interest in enantioselective growth of inorganic chiral materials has been reignited by demonstrations of unique properties such as optical activity (2, 3), response to magnetic fields (4) and chirality-induced spin selectivity (5, 6). Chiral semiconductors can also be used as circularly polarized photodetectors (7), which form a basis for optical quantum information technology (2), and topological homochiral crystals have shown enhanced catalytic activity for energy conversion and asymmetric synthesis (8).

While enantioselective synthesis of bioactive chiral molecules is routinely used for the production of pharmaceuticals (9, 10), the enantioselective synthesis of inorganic chiral materials remains challenging. Existing approaches rely on using chiral molecules in solution or on surfaces to template crystal growth (11–17). This is effective for colloidal nanoparticles but presents fundamental obstacles for electronic and photonic applications. Organic adsorbates introduce interfacial contamination that degrades charge transport and optical properties. The low temperatures and solvents required to preserve molecular templates are incompatible with the high-temperature vapor-phase deposition processes used to control atomic structure in semiconductor manufacturing. Moreover, the integration of chiral crystals into heterostructure devices requires precise control of interface structure. A method that transfers chirality directly from one crystal to another, without molecular intermediates, would overcome these limitations.

Epitaxy, the growth of one crystal (epilayer) on the surface of another crystal (substrate) with controlled atomic registry between them, is the standard method for producing high-quality crystalline thin films and heterostructures in the semiconductor industry. Crystallographic information from the substrate is transferred to the epilayer, including structure, morphology, orientation, and polarity (18). Attempts to use epitaxy to control chirality have been reported, including the growth of chiral metal surfaces on chiral planes of silicon (19), nanoribbons of $WS_2$ with chiral edge states on sapphire (20), and chiral planes of $ReS_2$ on asymmetric substrates with atomic steps (21, 22). In all these cases, however, only the terminations (surface or edges) of the grown crystals were chiral, while the bulk crystals themselves were achiral. The enantioselective growth of a chiral crystal via epitaxy (termed here 'chiral epitaxy' or 'chirotaxy') has not yet been demonstrated.

Achieving chiral epitaxy requires a substrate with a chiral surface and an epilayer with a chiral bulk structure. For the substrate we select $ReSe_2$, a two-dimensional (2D) van der Waals material belonging to the triclinic $P\bar{1}$ space group. It is achiral because it has inversion symmetry. However, this symmetry is broken at the surface, so the basal planes (001) and (00$\bar{1}$), corresponding to the two sides of a flake of the material, are enantiotopic (11), i.e. mirror-image surfaces of an achiral structure (Fig. 1B). For the epilayer, we select tellurium (Te), a chiral bulk crystal with a one-dimensional (1D) van der Waals structure, and a trigonal crystal lattice belonging to either of the two enantiomorphic space groups $P3_121$ (right-handed) or $P3_221$ (left-handed). Te was recently shown to be a Weyl semiconductor (23–27), and candidate for topological phase change transistors (28–30), and Te nanowires (NWs) exhibit chirality-induced spin selectivity (6), chirality-dependent thermoelectric (31), photovoltaic (6, 32), and photoresponse (33–35) properties. Aligned growth of Te NWs has been reported (34, 36), but without control over handedness.

Here, we demonstrate chiral epitaxy in the growth of Te NWs with controlled handedness on ReSe$_2$, with enantiomeric excess up to 73%. We study the mechanism of chiral epitaxy by *in situ* scanning and transmission electron microscopy, which reveal that the handedness is determined upon nucleation and does not change during growth. We develop a classical nucleation model, supported by *ab initio* density functional theory calculations and molecular dynamics simulations, that attributes the observed enantioselectivity to the interface energy difference between left-handed and right-handed Te nuclei on each ReSe$_2$ surface. Chiral epitaxy, demonstrated here in the Te/ReSe$_2$ system, establishes a potential route for the synthesis of inorganic chiral materials with controlled handedness, without the use of molecules and solvents, and at temperature and gas-phase conditions compatible with semiconductor manufacturing processes.

### Epitaxial growth of Te nanowires on ReSe$_2$

To obtain substrates with each of the two enantiotopic basal planes of ReSe$_2$ exposed, we sandwich a ReSe$_2$ crystal between two polydimethylsiloxane (PDMS) sheets and exfoliate it, then transfer each split half of the ReSe$_2$ crystal onto a silicon substrate (Fig. 1A schematic and 1B). This procedure ensures that one half flake has its (001) plane facing upward while the other has its (00$\bar{1}$) plane facing upward. Each plane can be macroscopically identified by the naked eye from its asymmetric parallelogram shape (see photograph at the center of Fig. 1A). We then grow Te NWs on these flakes via physical vapor transport, evaporating elemental Te at 400 °C and recrystallizing it onto the ReSe$_2$ substrates held at 220-240 °C. Scanning electron microscopy (SEM) images show that the Te NWs nucleated on the ReSe$_2$ surface grow aligned in one direction (Fig. 1D and 1E), forming dense arrays of parallel NWs with slightly varying widths and lengths (fig. S1 to S3).

To determine the atomic structure and epitaxial relationship at the Te/ReSe$_2$ interface, we prepared cross-sectional lamellae on several tens of aligned Te NWs (Fig 2A) using focused-ion beam milling (FIB). Lamellae were cut both perpendicular and parallel to the NW axis, enabling transmission electron microscopy (TEM) observations along two orthogonal directions. Most Te NWs have a trapezoidal cross-section shape (Fig. 1G). High-angle annular dark-field scanning transmission electron microscopy (HAADF-STEM) shows that all the Te NWs grow along the Te[0001] direction, which is the chiral axis of the Te crystal (Fig. 1F). Most Te NWs are single crystal, and the few occasional grain boundaries observed can be attributed to collision between NWs (fig. S12).

The crystal structure of Te can be intuitively described as a bundle of hexagonally packed covalent chains of divalent, tetragonal, sp$^3$-hybridized Te atoms. Because the chains are not straight, the most space-efficient and energetically stable way of packing them is in a helical configuration where they are either all left-handed or all right-handed (Fig. 1C). The Te atoms in the helical chains are located at the edges of a trigonal prism, appearing as triangles in the cross-sectional images (Fig. 1F). With respect to the ReSe$_2$, the Te NWs grow across the parallelogram Re$_4$ chains (Fig. 1B). Images in both the transverse (Fig. 1F) and longitudinal (fig. S4 to S6 and supplementary strain analysis) directions show an apparently commensurate interface, with the bottom Te atoms of the chains sitting above the Re atoms (*37*). The absence of an amorphous interlayer or secondary phases indicates that growth proceeds by direct epitaxy of the Te crystal in atomic contact with the ReSe$_2$ surface, which can enable enantioselective interactions.

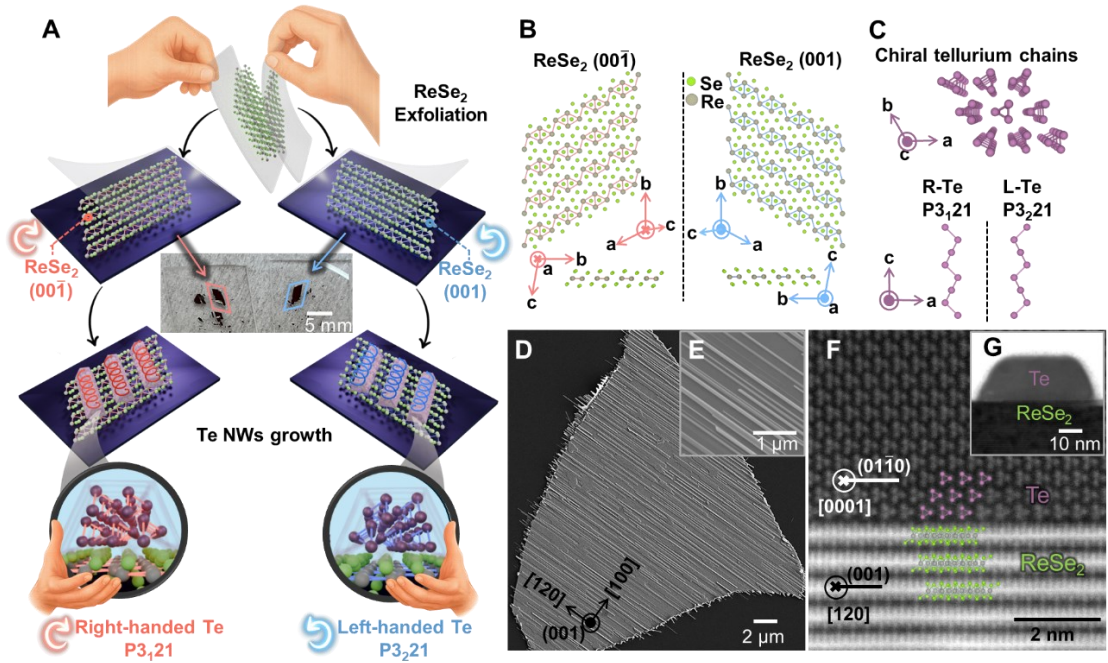

**Fig. 1. Chiral epitaxy of Te NWs on ReSe₂.** (**A**) Exfoliation of ReSe₂ onto silicon substrates with top (001) and (00$\bar{1}$) enantiotopic basal planes, followed by enantioselective epitaxial growth of chiral Te NWs on them. (**B**) Crystallographic projections of the enantiotopic (001) and (00$\bar{1}$) planes of ReSe₂ showing their parallelogram Re₄ chains, along **a**, where **a** = [100], **b** = [010], **c** = [001]. (**C**) Crystallographic projections along and across the chiral axis of Te, i.e. **c**, where **a** = [2$\bar{1}\bar{1}$0], **b** = [$\bar{1}$2$\bar{1}$0], **c** = [0001], (**D**) SEM top-view image of aligned Te NWs grown on a ReSe₂ flake. (**E**) High-magnification SEM of the aligned Te NWs grown on a ReSe₂ flake. (**F**) HAADF-STEM image across a Te NW on ReSe₂. (**G**) Low-magnification bright field-STEM image showing the trapezoidal cross-section shape of the Te NW.

### Enantioselective growth of chiral Te nanowires

The STEM cross-sections described above (Fig. 1F) reveal the atomic and epitaxial arrangement of each NW, but not the handedness of the Te crystal. When viewing the Te NWs in cross-section, the atomic chain projections that appear as triangles can be found with their mid-height vertices pointing all to the right (as in Fig. 1F and 2B₁) or all to the left (Fig. 2C₁). These are two different orientations of the Te crystal with respect to the ReSe₂ regardless of their handedness, however because the chains are viewed along their c-axis, Te [0001], both left- and right-handed Te chains appear the same. Distinguishing the two enantiomers requires viewing the crystal along a direction that breaks this degeneracy. To determine the handedness of the Te chains unambiguously, we tilt the sample by +20.6º or -20.6º around an axis perpendicular to the interface bringing the sample to the Te [2$\bar{1}\bar{1}\bar{6}$] or [$\bar{2}$116] zone axes (fig. S7). Upon these rotations, the appearances of left-handed and right-handed Te are clearly distinguishable, one looking like a series of U-shaped "happy" faces (Fig. 2B₂ and C₂), and the other looking as a series of upside-down-U-shaped "sad" faces (Fig. 2B₃ and C₃). A related approach using a different tilt axis and angle was previously used to determine the handedness of Te (*38*). We then determined the basal plane orientation of the ReSe₂ crystals, (001) or (00$\bar{1}$), by applying an analogous tilting method with ±30º tilts (Fig. 2 D₁ to D₃ and fig. S8).

Owing to the nearly perfect epitaxial alignment and high density of the Te NWs, we can cut across as many as 20-70 NWs in a single lamella (Fig. 2A), gathering a meaningful statistical ensemble of NWs per sample. The results show that up to 86% of the Te NWs growing on the (001) plane of ReSe$_2$ are left-handed (L-Te), whereas 14% are right-handed (R-Te). This represents a left-to-right enantiomeric excess, $EE = \frac{L-R}{L+R} \cdot 100\%$, of up to 73%. On the contrary, Te NWs on ReSe$_2$(00$\bar{1}$) are mostly R-Te, with $EE$ of up to -46%. $EE$ data obtained on seven samples is summarized in Fig. 2E, showing the consistency of the enantioselective and enantiospecific growth (Supplementary Materials, S1.9). Further experiments performed on suspended ReSe$_2$ flakes show that Te NWs grow on both sides of the same flake with similar enantioselectivity, as shown in supplementary materials (SM) and figs. S9 to S14. Materials and growth conditions must be chosen carefully to obtain high $EE$. On ReS$_2$, Te NWs did not show strong $EE$ under the conditions used for Te on ReSe$_2$ (fig. S15); furthermore, selenium (Se) deposited on both ReS$_2$ and ReSe$_2$ formed NWs that were shorter and less well aligned than Te NWs, making their handedness hard to analyze with good statistics (fig. S16). The model we describe below suggests that $EE$ will depend strongly on growth conditions, which must therefore be optimized for each material system to obtain high values of enantioselectivity.

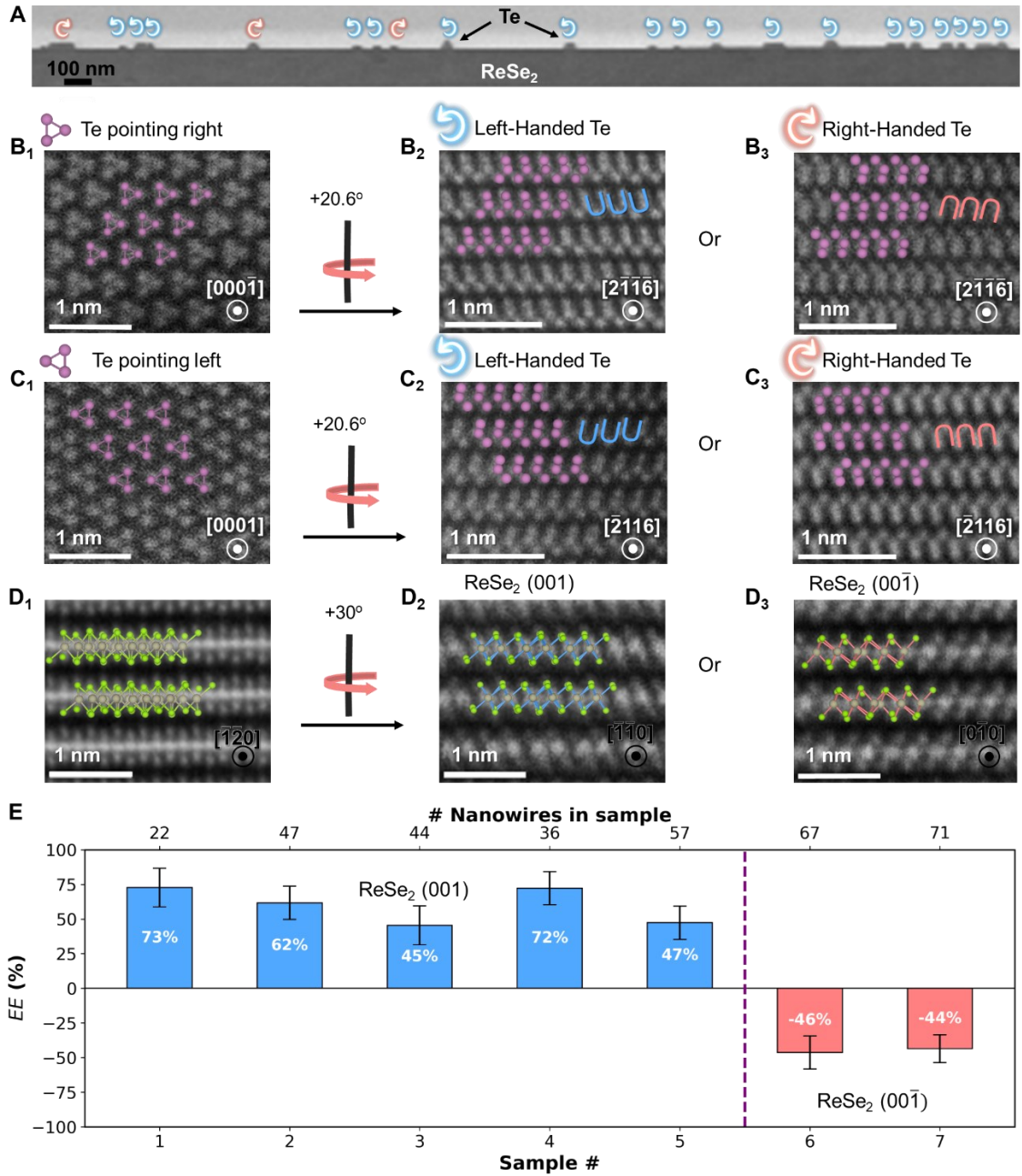

**Fig. 2. Enantioselective growth of right- and left-handed Te NWs on ReSe₂(00$\bar{1}$) and (001), respectively.** (**A**) Low-magnification dark-field image of the cross-section lamella. The NWs handedness is indicated by the right-handed red and left-handed blue arrows. (**B-D**) Show the fringes observed for the different combinations of Te handedness, Te orientation, ReSe₂ basal plane, and tilting direction. Te NWs were rotated to ±20.6º from the [0001] zone axis (**B** and **C**), and ReSe₂ was rotated by ±30º from the [120] zone axis (**D**). The zone axes for Te are marked in white, and for ReSe₂ in black. Atomic models are overlaid on the micrographs. The upright and upside-down "U" symbols indicate the happy and sad faces that characterize the left- and right-handed Te, respectively. (**E**) The *EE* recorded in each experiment (Sample #) is calculated as the excess of left-handed NWs. Negative values, indicating an excess of right-handed Te NWs, are

found on ReSe$_2$(00$\bar{1}$), whereas positive values, indicating an excess of left-handed Te NWs are found on ReSe$_2$(001), indicating enantioselective growth. The number of NWs in each lamella is shown in the upper chart axis. The vertical bars represent the *EE* standard deviation in each sample (see S1.9 in the SM). Samples were cut in central areas of the flakes, to avoid including NWs that could have nucleated at the flake edges by a different mechanism.

**Investigating the mechanism of chiral epitaxy by *in situ* electron microscopy**

The NW enantioselectivity demonstrated above could in principle arise at different stages of crystal formation. Handedness could be established during nucleation, when the first few atoms assemble on the substrate, or it could evolve during subsequent growth through restructuring or differences in growth rates. Distinguishing between these mechanisms requires direct observation of the growth process in real time. We therefore performed *in situ* experiments using both scanning and transmission electron microscopy (*in situ* SEM and *in situ* TEM) to track NW formation from nucleation to growth, to eventual coalescence into a continuous film.

For *in situ* SEM observations, we transferred exfoliated ReSe$_2$ flakes into an environmental SEM equipped with a heating stage and Te vapor source (SM, section S1.6). The growth conditions were tuned to obtain Te NWs in a vacuum environment and minimize the formation of other structures, such as non-planar NWs or nanotubes (fig. S17 to S19 and movies S1 to S3). The resulting SEM movies show that NWs nucleate first on the flake edges, cracks or step edges of the ReSe$_2$, and afterwards at the center of the ReSe$_2$ flake (Fig. 3A and fig. S19). Subsequently, the NWs elongate in both directions along their axis, rather than growing from one end only (Fig. 3A). This bidirectional growth, also seen in the TEM observations described below, is characteristic of a vapor-solid (VS) mechanism, in which adatoms attach along the full length of the crystal (*39–41*). As the growth continues, the NWs elongate and slowly widen until they coalesce to form a continuous Te 2D film on the ReSe$_2$ flake (Fig. 3A, fig. S19 and S2 and movies S4 and S5). Raising the substrate temperature causes the NWs to shrink and disappear as Te evaporates from the sample surface at a rate that increases rapidly with substrate temperature.

Kinetic analysis was performed for NWs longer than 1 μm, as shorter NWs could not be reliably measured due to SEM contrast and resolution limitations. At these NW dimensions we find that the NWs in the experiment, despite having different final dimensions and nucleation locations, have similar elongation and widening rates of $7 \cdot 10^{-4}$ μm s$^{-1}$ and $2 \cdot 10^{-5}$ μm s$^{-1}$, respectively. The lack of dependence of growth rate on NW size suggests that growth is limited by diffusion of Te adatoms across the substrate surface rather than by incorporation kinetics at the NW (Fig. 3B and C) (*40*).

To observe the atomic-scale processes governing chirality during growth, we repeated Te deposition experiments *in situ* in an environmental transmission electron microscope (ETEM) equipped with a molecular beam epitaxy source for Te deposition (*42*). ReSe$_2$ flakes were transferred onto *in situ* TEM heating grids to provide flat suspended membranes thin enough for imaging in transmission (SM, section S1.7). Figure 3D and movie S6-S7 show representative sequences of Te NWs growing on a suspended ReSe$_2$ flake. No catalyst is visible; instead the NW in movie S7 elongates and widens by stepwise addition of atomic layers onto its tip, sides, and top surface. Step flow is visible as new atomic layers nucleate on the NW surface and propagate toward its end.

The NW in this movie shows a change in tip faceting after it crosses between two regions of the ReSe$_2$ flake with different contrast, possibly a step edge or grain boundary such as those shown in figs. S1D and S2C.

Despite this change in tip faceting, no grain boundary is observed to form within the Te NW itself. In contrast, grain boundaries are readily detected in our experiments when NWs grow together and merge (see Supplementary Analysis, movie S6 and S8 and fig. S21. Since a change in handedness necessarily involves the formation of a boundary between domains of opposite chirality, the absence of detectable boundaries other than those due to coalescence suggests that the handedness established at nucleation is maintained, even when the NW encounters substrate defects.

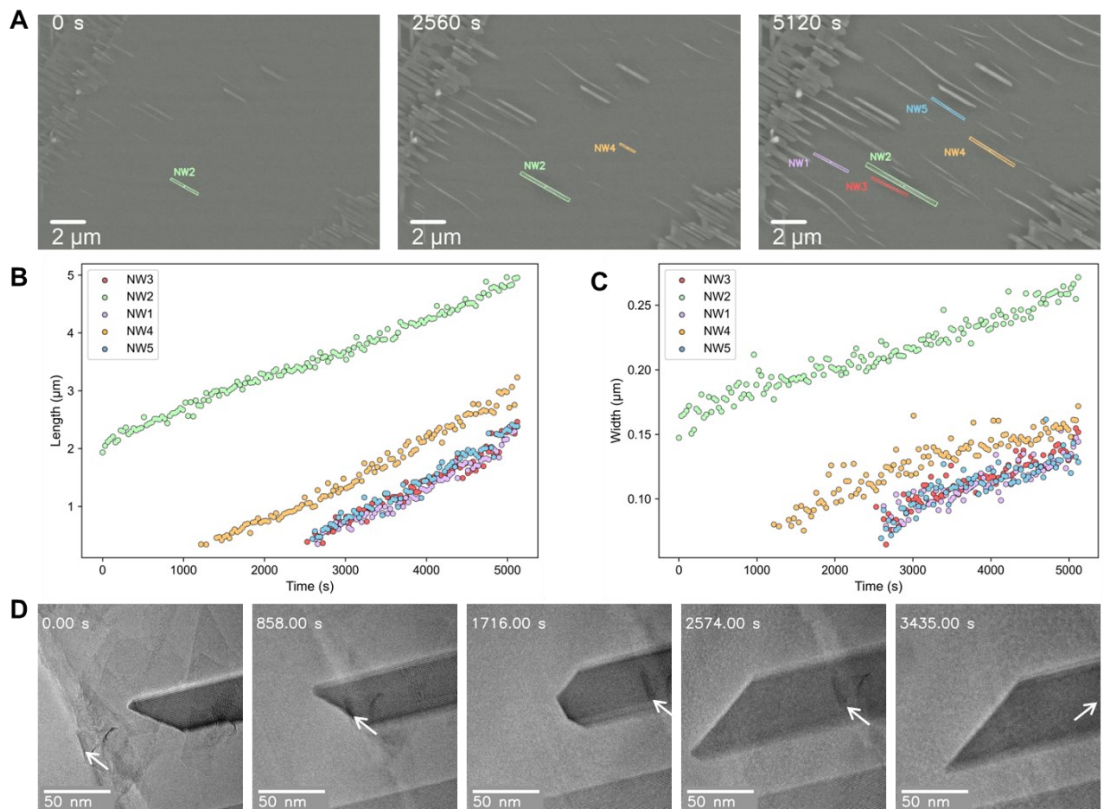

**Fig. 3. Real-time imaging of Te NWs growth on ReSe₂ using *in situ* electron microscopy.** Images sequence showing Te NW growth in the environmental SEM (**A**) with the time dependence of the length (**B**) and width (**C**) of individual NWs marked in respective colors in (**A**) (movies S4 and S5). (**D**) Image sequence obtained in the environmental TEM showing a Te NW during growth. The arrow marks the same location on each image (movie S7).

### Theoretical model of enantioselective growth

Based on the conclusion that handedness is determined at nucleation, we develop a simple model of enantioselective growth based on classical nucleation theory (*43*). Crystal formation begins with small clusters, or embryos, composed of atoms that arrive directly from the vapor or diffuse across the surface. Each embryo experiences two competing thermodynamic forces: adding atoms to the embryo releases free energy dependent on the embryo volume ($\Delta G_v$), favoring growth, but the size increase costs free energy due to the larger surface and interface areas ($\Delta G_s$), favoring dissolution (Fig. 4B). For small embryos the surface term dominates and the cluster is unstable, but beyond a critical radius ($r^*$) the volume term wins and the embryo becomes a stable nucleus that can grow into a full-fledged crystal, in this case a NW. Because left- and right- enantiomeric crystal embryos have different interface interactions with the chiral substrate that affect their total free energy, they should also have different critical sizes and activation energies ($\Delta G^*$) (Fig. 4A). A favorable interface, with low interfacial energy, reduces the total surface contribution and thereby lowers both the critical radius and the activation barrier to initiate faster nucleation of one Te enantiomer over the other.

Based on this model (SM, section S1.10) the enantiomeric excess $EE$ is given by Eq. 1, where $N_A$ is Avogadro's number, $M$ is the molar mass of Te, $\rho$ is the density of Te, $R$ is the universal gas constant, $\Delta H_{sub}$ is the sublimation enthalpy of Te, $\gamma_g$ is the surface energy of the nucleus-gas interface, $\Delta\gamma_{LR}$ is the interface energy difference between left- and right-handed Te nucleus on $ReSe_2(001)$ or $(00\bar{1})$, and $T$ is the sample temperature. $T_o$ is the critical deposition temperature at which the Te evaporation rate off the sample (due to atoms evaporating directly from nuclei or adatoms evaporating from terraces) is equal to the incorporation rate of Te onto the sample.

$$EE = tanh\left[-\frac{2\pi N_A M^2 \gamma_g^2}{RT\rho^2 \Delta H_{sub}^2} \cdot \Delta\gamma_{LR} \cdot \frac{T_0^2}{(T_0-T)^2}\right] \qquad \text{(Eq. 1)}$$

We calculate the interface energy difference $\Delta\gamma_{LR}$ by *ab initio* density functional theory (DFT, SM, section S1.12) to be -1.7×10$^{-3}$ eV Å$^{-2}$, which predicts that left-handed Te NWs should be thermodynamically more stable on $ReSe_2(001)$, and right-handed Te more stable on $ReSe_2(00\bar{1})$, as experimentally observed (Fig 2E). The $\Delta\gamma_{LR}$ value estimated from the experimental results using Eq.1 and literature values for $\Delta H_{sub}$ and $\gamma_g$ (*44, 45*) is -4×10$^{-3}$ eV Å$^{-2}$, which is quantitatively very close to the value calculated by DFT (SM, section S1.13).

Because the $T_o$-$T$ temperature difference appears in the denominator, $EE$ is predicted to increase dramatically as $T$ approaches the critical deposition temperature $T_0$. However, at that point the Te NWs would start to evaporate, so a 100% $EE$ cannot be fully achieved. The predicted $EE$ as a function of $T$ based on Eq. 1 is plotted in Fig. 4C, using the experimental values of $EE$ and $\Delta\gamma_{LR}$. The model assumes that handedness is fixed once a stable nucleus forms, but it leaves open the possibility that small embryos might interconvert between enantiomeric forms before reaching the critical size. A*b initio* molecular dynamics simulations show that embryos of one layer of Te chains undergo spontaneous transition to the more stable left-handed configuration when nucleating on $ReSe_2(001)$ (Fig. 4D, SM analysis), whereas two-layer embryos retain their initial handedness (SM, section S1.11).

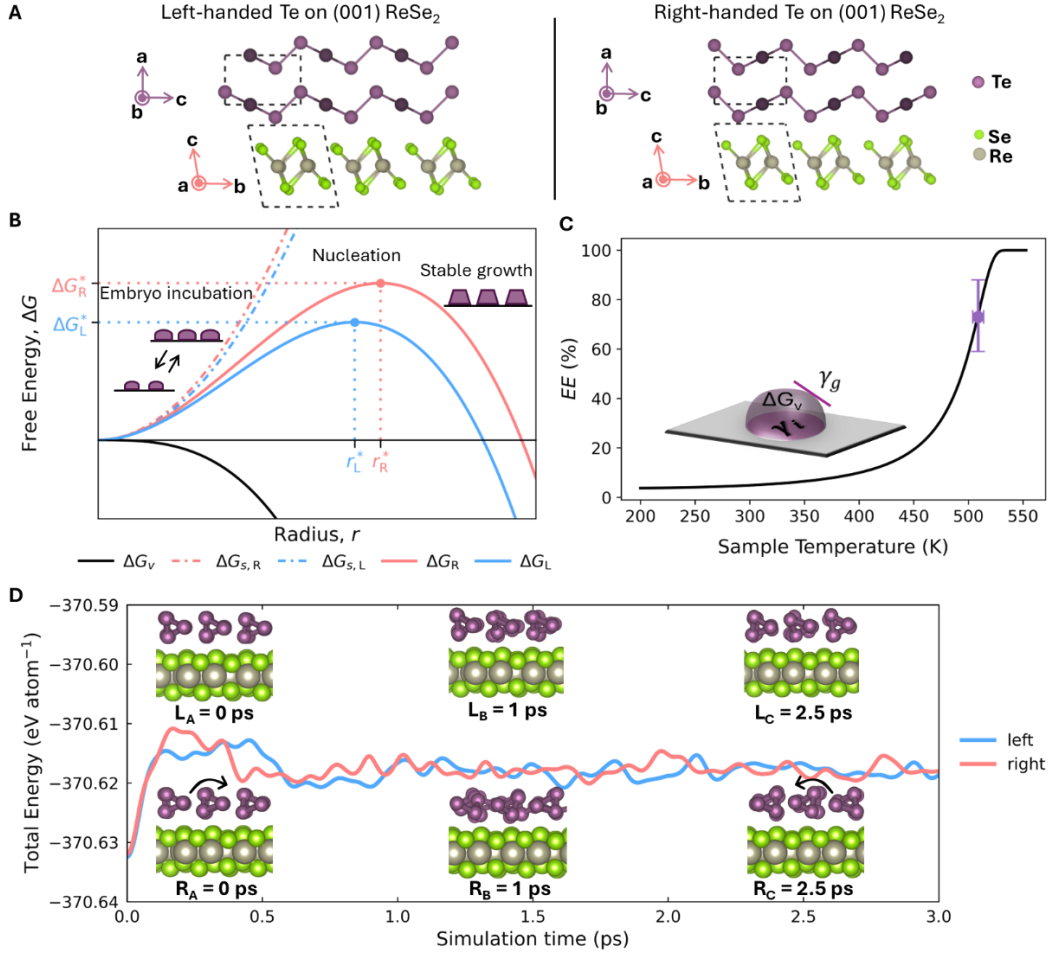

**Fig. 4. Suggested model for the enantioselective growth via chiral epitaxy. (A)** Difference between left- and right-handed tellurium NWs on ReSe$_2$(001), as seen across the c-axis of the NW. Te atoms are shown lighter when closer to the viewer. **(B)** Free energy profile for left- (blue) and right-handed (red) Te growth on ReSe$_2$(001) according to the theoretical model. The black solid line represents the volumetric Gibbs free energy, and the dashed lines represent the surface Gibbs free energy for left- (dashed blue) and right-handed (dashed red) nuclei. The solid blue and red graphs represent the Gibbs free energy of left- and right-handed nuclei, respectively. When the radius $r$ is below the critical value $r^*$, nuclei are unstable and dissolve, whereas above $r^*$, they grow stably. **(C)** Enantiomeric excess (*EE*) predicted by the theoretical model as a function of sample temperature $T$ for sample #1. The experimental *EE* value is marked purple, with error bars. The inset sketch shows the initial Te nucleus droplet with the corresponding volumetric free Gibbs energy of the droplet ($\Delta G_v$), the surface energy of the nucleus-gas interface ($\gamma_g$) and the surface energy of the nucleus-substrate interface ($\gamma_L$ or $\gamma_R$). **(D)** Molecular dynamics simulations for the total surface energy of left-handed (blue) and right-handed (red) Te on ReSe$_2$(001). The simulation shows that right-handed Te nucleating on (001) basal plane of ReSe$_2$ can change handedness after 2.5 ps.

## Conclusions

We have demonstrated chiral epitaxy (or 'chirotaxy'): the enantioselective growth of intrinsically chiral crystals from the vapor phase on a chiral substrate surface. We show that Te

NWs grown on the low-symmetry $ReSe_2$ surface exhibit enantiomeric excesses up to 73%, with the (001) surface favoring left-handed Te and the $(00\bar{1})$ surface favoring right-handed Te. *In situ* SEM and TEM reveal that handedness is established at nucleation and preserved throughout growth. A classical nucleation model, supported by density functional theory and molecular dynamics calculations, attributes the selectivity to the interface energy difference between Te enantiomers and the chiral substrate. This process differs from existing approaches to enantioselective crystal synthesis, such as molecular templating, which rely on solution-phase growth. Chirotaxy utilizes vapor-solid growth and operates at elevated temperatures, produces atomically clean interfaces, and requires no solvents or organic additives. This aligns with established semiconductor manufacturing processes, suggesting a pathway toward integration of homochiral materials into electronic and quantum devices, such as spin valves for spintronics and circular polarized photodetectors for quantum computing. For Te nanowires, the preservation of chirality after nucleation implies that chirotaxy could be seeded at specific locations and growth can then extend laterally onto other parts of the substrate, including amorphous regions, increasing the range of possible device applications.

The $Te/ReSe_2$ system provides an effective platform for demonstrating chirotaxy since the aligned nanowires enable statistical sampling across large ensembles. However, chirotaxy can be generalized beyond nanowires and 2D substrates. Our model suggests that any pairing of a substrate that has a chiral surface and an epilayer that has a chiral bulk structure could exhibit chirotaxy, if the interface energies are sufficiently different between enantiomers. Chirotaxy could thus be extended to many other intrinsically chiral materials, such as α-HgS (*12*), $TeO_2$ (*46*), chiral perovskites (*47*) and chiral transition-metal silicides and germanides (*8*) as well as to other chiral substrates like Si(643) (*19*), α-quartz or A-plane sapphire (see Supplementary Analysis). Chirotaxy establishes a vapor-phase, molecule-free route to homochiral crystals, adding handedness to the structural properties that epitaxy can control and opening a pathway toward integration of chiral materials into emerging device architectures.

**Acknowledgments:**

We thank the Weizmann Institute of Science AI-Hub for their guidance for the *in situ* SEM movie analysis, and Meir Lahav, Assaf Ben-Moshe and Lia Addadi for their critical comments on earlier versions of this manuscript. We acknowledge the Centre Interdisciplinaire de Microscopie Electronique de l'X in Palaiseau for enabling to carry out the ETEM experiments using the NanoMAX instrument, and the Moskowitz Center for Nano and Bio-Nano Imaging.

**Funding:**

US-Israel Binational Science Foundation (BSF), grant No. 2020096 (EJ, FMR)

Israel Science Foundation (ISF), grant No 1045/23 (EJ)

"Equipements de Recherche Mutualisés" ERM 2021 grant (FP)

Ariane de Rothschild Women Doctoral Program (NRI)

Hugh Hampton Young Memorial Fund Fellowship at MIT (KReidy)

Aryeh and Mintzi Katzman Professorial Chair and the Helen and Martin Kimmel Award for Innovative Investigation (LK)

Financial support of the Heinmann Chair in Physical Chemistry (OHod)

Drake Family Professorial Chair of Nanotechnology (EJ)

**Author contributions:**

Conceptualization: NRI, KReidy, MLG, FMR, EJ

Project Direction: EJ and FMR

Samples preparation and characterization for *ex situ* experiments: NRI, MLG, AA

*In situ* SEM experiments: NRI, XMS, IKA

Lamellae preparation: NRI, KRechav

STEM characterization: NRI, MLG, OB, LH

Sample preparation for *in situ* TEM experiments: KReidy, PAM

*In situ* TEM experiments: CW, KReidy, PAM, NRI, FP, EJ, FMR

*Ab initio* DFT calculations: JG, RT, OHellman, NRI, EJ, OHod, LK, DSG, PASA

*Ab initio* molecular dynamics simulations: JGQ, RT, NRI, EJ, DSG, PASA

Development of the theoretical model for enantioselective growth: SJ, NRI, EJ

Supervision: EJ, FMR, FP, PASA, DG, LK, OHod

Writing – original draft: NRI, KReidy, EJ, FMR

Writing – review & editing: All contributing authors

**Competing interests:** Authors declare that they have no competing interests

**Data, code, and materials availability:** All data are available in the main text or the supplementary materials. Code used for movie analysis is available upon request from authors.

**Supplementary Materials**

**Other Supplementary Materials for this manuscript include the following:**

### S1 Materials and Methods

### S1.1 Exfoliation of the ReX$_2$ (X = S, Se) flakes

ReX$_2$ bulk crystals were purchased from HQ Graphene company (Netherlands). A single bulk crystal was sandwiched between two PDMS sheets and exfoliated carefully several times. Silicon substrates with either native oxide or 295 nm thermal oxide were sonicated in acetone and IPA for 10 minutes. Then, the substrates were heated at 250 °C for 10 minutes to remove moisture and organic residues. After a short cooling, the ReX$_2$ flakes were transferred to the silicon substrates. The substrate was washed for 1 min with acetone and IPA and then dried with N$_2$ flow to remove tape residue. Then, the samples were heated again at 250 °C for 7 minutes for adhesion of the flakes to the substrate.

### S1.2 Chemical vapor deposition of tellurium nanowires

The growth of tellurium (Te) nanowires (NWs) on ReX$_2$, is done in the CVD system. Te source powders (99.999%, Holland Moran) are placed in the hot zone of a three-zone tube furnace (Blue M), while the samples were placed downstream in the colder zone of the furnace. Before growth, the quartz tube is purged to avoid oxygen and moisture residues. During growth, a mixture of 70 sccm of N$_2$ (99.999%, Gordon Gas) and 10 sccm H$_2$ (99.99995%, Parker Dominic Hunter H$_2$-generator) gases is used to carry the source vapor. The Te powder is kept at 400 °C, and the substrate is placed in the second heating zone at 200-250 °C. Pressure is set to 35 mbar. After 10-15 min of growth, the furnace is quickly moved away to allow rapid cooling of the sample and source to room temperature.

### S1.3 Chemical vapor deposition of selenium NWs

Selenium (Se) NWs growth was done in a CVD system. Two approaches were used to grow the Se NWs, based on previously reported synthesis conditions for producing Se NWs were either hydrogen or graphite are used to form a reducing synthetic environment. The first approach, Se powder (99.99%, Sigma Aldrich) was placed in the hot zone of a three-zone furnace (Blue M), and the sample was placed downstream in the colder zone. The quartz tube was purged to avoid oxygen and moisture contamination. During growth, a mixture of 50 sccm of N$_2$ (99.999%, Gordon Gas) and 5 sccm H$_2$ (99.99995%, Parker Dominic Hunter H$_2$-generator) gases was used to carry the source vapor. During synthesis the Se powder was kept at 350-210 ℃, while the substrate was kept at 250-100 ℃. Pressure was set to 35 mbar. After 20 minutes of synthesis, the furnace was quickly moved away to allow rapid cooling. The second approach kept the same growth conditions but used a mixture of Se and graphite powder (Sigma Aldrich, 99.99%) and no $H_2$ flow.

### S1.4 Structural and elemental characterization of the Te/ReX$_2$ heterostructures

(i) *Optical measurements and scanning electron microscopy (SEM).* The growth results are assessed using an optical microscope (Olympus BX-51). High-resolution imaging of the obtained nanostructures and devices is done mainly by field-emission SEM (SIGMA and ULTRA 5 Zeiss, voltages 1-5 kV).

(ii) *Atomic force microscopy (AFM).* Determining the height and surface morphology of the mixed-dimensional heterostructures was done using atomic force microscopy (AFM, Veeco, Multimode Nanoscope 7.30). Images were taken using tapping mode in open air at 70 kHz (RTESP7) etched Si tips (Nanoprobes), followed by image processing using Gwydion software.

(iii) *Transmission electron microscopy (TEM) and energy-dispersive X-ray spectroscopy (EDS) characterizations.* Aberration-corrected transmission electron microscopy was employed to investigate the crystallographic structure, orientation, and epitaxial relationships between the NWs and the 2D layered material. Imaging and analytical measurements were performed on a double aberration-corrected Themis Z microscope (Thermo Fisher Scientific Electron Microscopy Solutions, Hillsboro, USA) equipped with a high-brightness field emission gun operated at 200 kV. High-angle annular dark-field scanning transmission electron microscopy images were acquired using a Fischione Model 3000 detector with a semi-convergence angle of 21 mrad, a probe current of ~50 pA, and an inner collection angle of 60 mrad. Elemental analysis of the mixed-dimensional heterostructure was performed using energy-dispersive X-ray spectroscopy with hyperspectral data collected using a Super-X G2 four-segment silicon drift detector with a probe semi-convergence angle of 21 mrad and a beam current of ~200 pA. The TEM samples were prepared using a focused ion beam (FIB, Helios 600 FIB/SEM Dual Beam Microscope, Thermo Fisher Scientific) for cutting a thin lamella across and along the supported ReX$_2$ flakes with tens of NWs. A thick conductive carbon coating layer was deposited to protect the nanostructures using the Edwards coater and the FIB microscope. Using the ion beam, the lamella's sides were carved and then thinned and polished using Ga$^+$ ions after mounting it on a TEM grid.

### S1.5 Handedness determination for the Te NWs

Handedness determination is carried out by performing tilting series imaging of the Te NWs, where the entire lamella is tilted to the same tilt angle and the handedness of each NW is determined as described in fig. S5, S7 and S14. The atomic projections were visualized using VESTA software (*48*). Overall, we analyzed three different samples orientations: cross-sectional lamella, longitudinal lamella (cut along the NWs), and a plan-view orientation, where Te NWs were grown on ReSe$_2$ flakes suspended on a holey SiN TEM membrane. For each orientation, a different rotation angle was used to determine the handedness.

### S1.6 Handedness determining the basal plane of ReX$_2$

The determination of the basal plane handedness of ReX$_2$, i.e. (001) or (00$\bar{1}$), was done by performing tilt series imaging of the ReX$_2$. Two different tilting angles were used when analyzing the cross-sectional lamella and the longitudinal lamella, as described in fig. S8 and fig. S6. Determining the orientation of the plan-view samples, where ReSe$_2$ basal plane is suspended on a TEM membrane, was based on analysis of the FFT image of the ReSe$_2$. Because the two basal planes are mirror images of one another, so are the FFT images of the crystal.

### S1.7 *In situ* SEM growth experiments

(i) *In situ SEM growth experiments of Te NWs on ReSe$_2$.* For this purpose, we used the ESEM (Quattro S, Thermo Fisher Scientific) equipped with a gas secondary electrons detector (GSED) and a 1400ºC heating stage. An MgO crucible filled with ~0.05 gr of Te powder (American Elements, 99.999%) was placed inside the heating cavity. A cylindrical metal cap (stainless steel or Ultra-Corrosion-Resistant Grade 2 Titanium, McMaster-Carr) was used as a spacer between the powder source and ReSe$_2$ on a silicon substrate. To further control the temperature gradient, we placed a copper bridge over the spacer and the fibrous heat insulation. Further explanations can be found here (*40*). The heating stage was heated to 240-350 ºC, the chamber pressure was 100-200 Pa and the imaging voltage was set to 20 kV.

(ii) *Image analysis for in situ SEM growth videos of Te NWs on ReSe$_2$.* All the *in situ* SEM movies were aligned using Fiji/ Image J (National Institute of Health, Maryland, USA) "Linear Stack Alignment with SIFT" plugin. For the movie used for data analysis of the growth rates of the Te NWs, further manual alignment was done using Fiji "TrakEM2" plugin. The image contrast was adjusted manually in Fiji. Ten representative frames, evenly distributed across the full duration of each movie, were selected for training a pixel classification model in ilastik (*49*), using two classes, Te NWs and ReSe$_2$. The trained model was then applied to the entire image sequence to generate binary segmentation images, which were exported and compiled into a new 2D image stack. Post-processing of the segmentation was performed in Fiji, where small artifacts were removed using the Analyze Particles plugin, and holes within the segmentations were filled using the Binary plugin. A custom Python script was then used to contour the segmented NWs and track their size throughout the movie. Finally, the tracking contours were overlaid onto the aligned original movie frames.

### S1.8 *In situ* TEM growth experiments

(i) *Suspended ReSe$_2$ TEM grid preparation.* Thermally grown 90 nm SiO$_2$/Si wafers were pre-treated with oxygen plasma and ReSe$_2$ (HQ Graphene, Groningen, Netherlands) was mechanically exfoliated onto them using the conventional Scotch tape method (magic 3MTM scotch tape). Flakes of few layers thickness were then identified by their contrast in optical microscopy. Identified flakes were transferred with a polymer-assisted transfer method, using cellulose acetate butyrate (CAB) as handle, as described previously (*50, 51*). Due to the hydrophilic nature of the CAB on the hydrophilic SiO$_2$, an added droplet of deionized water could intercalate to readily separate the CAB and ReSe$_2$ from the SiO$_2$ substrate. We then transferred this CAB handle to commercial TEM MEMS chips (Protochips) featuring a silicon carbide or silicon nitride membrane with 10-μm holes. The chips were baked at a temperature of ~100 °C for 2 minutes and subsequently the polymer was dissolved in acetone for ~15 min. Finally, the chips were dipped in isopropanol and dried using a critical point dryer.

(ii) *In situ TEM growth experiments.* High-resolution *in situ* growth was performed using a modified environmental transmission electron microscope (Titan ETEM 60-300ST) at a working voltage of 300 kV. The Te flux was supplied by a full PBN thermal cracker effusion cell (MBE

Komponenten), operating under a base vacuum of $5 \times 10^{-8}$ mbar. A custom-made sample holder, fabricated by Protochips, was used; it was designed to provide a direct line of sight between the effusion cell and the chip surface. Videos were recorded using a CCD US1000 camera (Gatan) at a frame rate of four frame per second with image size of 1k x 1k.

### S1.9 Error estimations for the enantiomeric excess

The enantiomeric excess ($EE$) for each sample was calculated from the relative number of left- and right-handed Te NWs, assuming a binomial distribution where the probability of each handedness is determined by their respective counts. The standard error ($SE$) was derived from this distribution and applied to the handedness counts. Using the SE-adjusted counts, the corresponding range of $EE$ values was recalculated. The error bars in the $EE$ plots reflect the minimum and maximum $EE$ values obtained from these SE-corrected counts.

### S1.10 Development of the theoretical model describing the enantioselective growth

The enantioselective growth of Te NWs is based on a classical nucleation theory, combined with the Clausius-Clapeyron equation to relate the vapor pressure to the temperature. First, we describe the Gibbs free energy for the formation of a nucleus on a substrate $\Delta G$, considering a volumetric component $\Delta G_v$ for the gas to solid transition, and a superficial component $\Delta G_s$ for the interfaces of the nucleus with the gas and the substrate (Eq. S1). For simplicity, we choose a hemispherical nucleus of radius $r$. Then, $\Delta G$ is given by Eq. S2, where $g_v$ is the volume energy for the vapor-to-solid transition, $\gamma_1$ is the surface energy at the nucleus-substrate interface and $\gamma_g$ is the surface energy at the nucleus-gas interface.

$$\Delta G = \Delta G_v + \Delta G_s \qquad \text{(Eq. S1)}$$

$$\Delta G = -\frac{2\pi r^3}{3} g_v + \pi r^2 (2\gamma_g + \gamma_1) \qquad \text{(Eq. S2)}$$

The critical radius $r^*$ refers to the smallest size a nucleus must reach for continuous growth to occur. When the nucleus radius is smaller than $r^*$, the nucleus will dissolve. The critical radius is derived by taking the first derivative of Eq. S1 and finding the maximum point.

$$\frac{d(\Delta G)}{dr} = -2\pi r^2 g_v + 2\pi r (2\gamma_g + \gamma_1) = 0 \qquad \text{(Eq. S3)}$$

$$\rightarrow r^* = 0 \; ; r^* = \frac{2\gamma_g + \gamma_1}{g_v}$$

Substituting $r^*$ into Eq. S1 yields the critical Gibbs free energy $\Delta G^*$, which is the activation energy of nucleation

$$\Delta G^* = \frac{\pi (2\gamma_g + \gamma_1)^3}{3g_v^2} \qquad \text{(Eq. S4)}$$

The enantiomeric excess ($EE$) is defined as the difference in concentration of one enantiomer over the other, in this case left-handed over right-handed, as described in Eq. S4.

$$EE = \frac{[L]-[R]}{[L]+[R]} \qquad \text{(Eq. S5)}$$

Based on our experimental observation that the handedness of the Te NWs does not change during growth, we assume that the handedness is determined at the nucleation stage. Using the Arrhenius equation, we assume that the concentration of right- and left-handed nuclei is proportional to the rate constant $k$, which is defined separately for right- and left-handed nuclei. Substituting the critical Gibbs free energy ($\Delta G^*$) in Arrhenius equation we can define $k_R$ and $k_L$ for right- and left-handed nuclei as described:

$$k_L = A e^{-\frac{\pi(2\gamma_g+\gamma_L)^3}{3g_v^2RT}} \;\; ; \; k_R = A e^{-\frac{\pi(2\gamma_g+\gamma_R)^3}{3g_v^2RT}} \qquad \text{(Eq. S6)}$$

Substituting the rate constants in Eq. S4 and the using the identity $\tanh(x) = \frac{e^x-e^{-x}}{e^x+e^x}$, we can describe the enantiomeric excess as follows:

$$EE = \frac{[L]-[R]}{[L]+[R]} = \tanh\left\{\frac{\pi}{6RTg_v^2}\left[(2\gamma_g+\gamma_R)^3-(2\gamma_g+\gamma_L)^3\right]\right\} \qquad \text{(Eq. S7)}$$

The volume free energy $g_v$ of Te is given by

$$g_v = \frac{\rho RT}{M}\ln\left(\frac{P_{Te}}{P_{Te}^v}\right) \qquad \text{(Eq. S8)}$$

where $\rho$ is the density of Te, $M$ is the molar mass of Te, $P_{Te}$ is the Te pressure around the sample and $P_{Te}^v$ is the vapor pressure of Te. $P_{Te}$ is not known, but can be inferred from the temperatures of the source and the sample using the Clausius-Clapeyron equation (Eq. S9).

$$\frac{dP}{dT} = \frac{P\Delta H_{sub}}{T^2R} \;\rightarrow\; \ln\left(\frac{P_{Te}}{P_{Te}^v}\right) = -\frac{\Delta H_{sub}}{R}\left(\frac{1}{T_2}-\frac{1}{T_1}\right) \qquad \text{(Eq. S9)}$$

Here, $T_1$ and $T_2$ would be the temperatures of the source and the sample, respectively, assuming that the vapor and the solids in both regions are found in thermodynamic equilibrium. In practice, however, there is a constant flow of gas and the system if found in a pseudo-equilibrium steady state that is not necessarily a thermodynamic equilibrium. We can solve this by assuming $T_1$ to be an effective source temperature, $T_0$, corresponding to the critical sample temperature at which the solid is deposited at the same rate as it evaporates, while $T_2$ is the actual sample temperature $T$. It is assumed that the transitions from vapor to solid and *vice versa* take place via series of intermediate species, such Te clusters in the vapor and Te adatoms on the surface, which form and react faster than the other steps of surface diffusion and crystal nucleation and growth. Under these assumptions, when $T = T_0$, equation Eq. S8 nullifies, i.e. $g_v = 0$, corresponding to the situation where NWs neither grow nor evaporate, by which $T_0$ is defined.

Substituting Eq. S9 into the volume free energy, squaring, and using the above definitions of $T_1 = T_0$ and $T_2 = T$, we obtain

$$g_v^2 = \frac{\rho^2R^2T^2}{M^2N_A}\left[\frac{\Delta H_{sub}^2}{R^2}\left(\frac{T_0-T}{T_0T}\right)^2\right] \qquad \text{(Eq. S10)}$$

Then, substituting Eq. S10 into Eq. S7:

$$EE = tanh\left\{\frac{M^2 N_A \pi R^2 T_0^2 T^2}{6\rho^2 R^3 T^3 \Delta H_{sub}^2 (T_0 - T)^2}\left[(2\gamma_g + \gamma_R)^3 - (2\gamma_g + \gamma_L)^3\right]\right\} \quad \text{(Eq. S11)}$$

Because a solid-gas interface energy is generally much higher than that of a solid-solid interface, we can assume $\gamma_L, \gamma_R \ll \gamma_g$, which simplifies the cube difference to the following:

$$(2\gamma_g + \gamma_R)^3 - (2\gamma_g + \gamma_L)^3 \approx -12\gamma_g^2 \Delta\gamma_{LR} \quad \text{(Eq. S12)}$$

where $\Delta\gamma_{LR} = \gamma_L - \gamma_R$ is the surface energy difference between left- and right-handed enantiomers of the solid and the substrate.

The enantiomeric excess $EE$ can then be described as Eq. S13 (which appears as Eq, 1 in the main text):

$$EE = tanh\left[-\frac{2\pi N_A M^2 \gamma_g^2}{RT\rho^2 \Delta H_{sub}^2} \cdot \Delta\gamma_{LR} \cdot \frac{T_0^2}{(T_0 - T)^2}\right] \quad \text{(Eq. S13)}$$

The critical parameters in this equation are the difference between the sample temperature and critical temperature, $T_0 - T$, and the surface energy difference, $\Delta\gamma_{LR}$, which can determine whether a certain combination of chiral epilayer and substrate materials could yield high or low enantiomeric excess values.

### S1.11 Calculation of the surface energy difference between left- and right-handed Te on ReSe$_2$ surfaces and molecular dynamics simulations of the Te chiral stability

We carried out a comprehensive study on the structural stability and total energy configurations of a single-layer ReSe$_2$ as substrate and top (bottom) layers of Te wires with left and right chirality deposited on the (001) and (00$\bar{1}$) basal planes of ReSe$_2$. The supercells of the Te-ReSe$_2$ heterostructure used for the calculations were generated based on the epitaxial relations experimentally observed by TEM. Figure S22 shows the structural model employed: a ($4 \times 4$) ReSe$_2$ supercell accommodating one or two layers of six tellurium chains per supercell. Periodic boundary conditions were applied in all directions; therefore, a vacuum spacing of ~18 Å was introduced along the out-of-plane (z) direction to suppress spurious interactions between periodic images.

The calculations were performed using *ab initio* density functional theory (DFT) as implemented in the SIESTA software (*52, 53*). GGA-PBE type exchange and correlation functionals were used to describe the electron-electron interactions (*54*). To describe the electron-ion interactions, the Troullier-Martins set of norm-conserving pseudopotentials in the Kleinman-Bylander form was used, with a doubly polarized $\zeta$ basis set.(*55*) The structural relaxations were performed using a real space mesh cut-off of 350 Ry. An electronic temperature of 300 K was used to smooth out the Fermi step function. The convergence criterion for the density matrix was set at $10^{-4}$. During the relaxation process, the Conjugate Gradient (CG) method was used for ionic optimization until the Hellmann-Feynman forces (*56*) were below 0.01 eVÅ$^{-1}$. The Monkhorst-Pack (*57*) k-sampling grid set was 1×1×1. The results of the DFT calculations are presented in Table S3. The table shows that the lowest energy minimum point is achieved for left-handed Te on (001) ReSe$_2$ and right-handed Te on ReSe$_2$ (00$\bar{1}$).

Interestingly, there is a slightly higher local energy minimum for the opposite configuration, i.e. right-handed Te on (001) ReSe$_2$ and left-handed Te on ReSe$_2$ (00$\bar{1}$), which could explain why the experimental $EE$ values are lower than expected from the calculated $\Delta\gamma_{LR}$ for the global minima for each configuration.

Beyond the zero Kelvin temperature total energy calculations used to quantify the relative energetic ordering of the different configurations of Te, we assessed their dynamic stability on ReSe$_2$ (001) and (00$\bar{1}$) surfaces, explicitly considering the presence of a second layer of Te chains. To this end, we performed *ab initio* molecular dynamics simulations (*52, 53*), which enable us to probe finite-temperature structural stability as well as possible changes in chirality and interlayer locking effects induced by the second Te layer. All simulations were carried out at room temperature, T = 300 K, within the canonical NVT ensemble using a Nosé thermostat (*58*), for a total simulation time of 10 ps with a time step of 1 fs.

### S1.12 Additional DFT calculation of the interface energy difference between right- and left-handed Te nuclei on ReSe$_2$ (001)

For an additional approach to calculate the surface energy difference between right- and left-handed Te nuclei on the basal plane of ReSe$_2$, $\Delta\gamma_{LR}$, we consider two thermodynamic states: one where the Te nucleus is in contact with the ReSe$_2$ surface, and the second where the two are separated with no interfacial interaction (fig. S23). The interfacial surface energy is defined as $\gamma_{X\text{-}ReSe_2}$, where X refers to either the right- (R) or left-handed (L) nucleus. Each facet of the crystal contributes its own surface energy, as shown schematically in fig. S23. Since the surface energy is treated as a state function, we define the energy difference between the contact and separated configurations as $\Delta E_{Contact-Apart}^{X}$.

$$\Delta E_{Contact-Apart}^{R} = \gamma_{R-ReSe_2} - (\gamma_{ReSe_2-up} - \gamma_{R-down}) \qquad \text{(Eq. S14)}$$

$$\Delta E_{Contact-Apart}^{L} = \gamma_{L-ReSe_2} - (\gamma_{ReSe_2-up} - \gamma_{L-down}) \qquad \text{(Eq. S15)}$$

Assuming that the surface energies of the standalone Te nuclei are the same for both enantiomers, the surface energy difference between the two enantiomers can naively be defined as the difference between the contact and separated configurations of each enantiomer:

$$\Delta\gamma_{LR} = \gamma_{L-ReSe_2} - \gamma_{R-ReSe_2} \qquad \text{(Eq. S16)}$$

However, that fails to take into account that the amount of strain for the right- and left-handed configurations is different. Therefore, we take instead the difference in contact energies:

$$\Delta\gamma_{LR} = \Delta E_{Contact-Apart}^{L} - \Delta E_{Contact-Apart}^{R} \qquad \text{(Eq. S17)}$$

Alignment of the Te atoms with respect to the ReSe$_2$ was obtained based on the epitaxial relations revealed from the TEM results. Four combinations, for right- and left-handed Te and for (001) and (00$\bar{1}$) ReSe$_2$ basal planes, were constructed. The DFT calculations were performed within the projector augment wave (PAW) formalism as implemented in VASP (*59–62*), using the PBE exchange-correlation functional (*54*). The plane wave cutoff was set to 330 eV and a 2x2x1 k-point mesh was used for Brillouin zone integration. To keep the size of the supercell computationally manageable, the ReSe$_2$-Te interface is slightly strained, with different residual stresses. In practice, a cell containing 80 Re atoms, 160 Se atoms, and 108 Te atoms was used.

Because the strain energy scales with the volume and the surface energy scales with the area, the residual strain may dominate the energy differences. Therefore, we estimate the surface energy by keeping the strain constant and separating the Te from the ReSe$_2$, to obtain configurations of constant strain such that the extraneous strain energy is largely canceled out. The calculated surface energy differences, as stated in Eq. S17 was then found to be -0.27 eV.  To obtain an energy normalized by the surface energy of the interface, we obtained:

$$\Delta\gamma_{LR} = \frac{-3.2351}{156.5526} - \frac{-2.9682}{156.5518} = -1.7 \cdot 10^{-3} \frac{eV}{\text{Å}^2} \qquad \text{(Eq. S18)}$$

Note the slight difference in the area used for normalization, which indicates the above-mentioned residual strain difference.

We note that although our calculations are based on the PBE GGA functional, which inherently does not account for long-range dispersion interactions, we obtain good agreement with the experimental structural parameters. Likely this involves at least some cancellation of errors. Nonetheless, given the appropriate geometry of the system, our attention is focused on evaluating the differences between left- and right-handed structures. As van der Waals interactions for both systems are expected to be similar, we can estimate these differences without explicitly accounting for the van der Waals energy (*63*).

### S1.13 Estimation of the interface energy difference from the experimental results

In order to validate the our theoretical model and calculations, we compare the  interface energy difference $\Delta\gamma_{LR}$ calculated by DFT with the one calculated from the experimental results using the theoretical model (S1.10). For this, we invert Eq. S13 to Eq. S19, and plug in all the parameters either from the experimental conditions and results or from literature values. The molar mass, $M$, of tellurium is 52 g mol$^{-1}$, the gas constant, $R$, is 8.314 J mol$^{-1}$ K$^{-1}$, the density of tellurium, $\rho$, is 6.24 g cm$^{-3}$ (*45*), the surface energy of tellurium, $\gamma_g$ at 450 ºC is 186 mN m$^{-1}$ (*45*), the sublimation enthalpy, $\Delta H_{sub}$, is 156.6 kJ mol$^{-1}$ (*44*) and the Avogadro's constant, $N_A$, is $6.022\cdot10^{23}$ mol$^{-1}$. Using the experimental values from sample #1, the sample temperature $T$ was 509 K (236 ºC) and the *EE* value is 0.73 (73%). The critical temperature $T_0$ was roughly estimated in a series of experiments where $T$ was raised until no Te NWs grew because they started to evaporate at the same rate as they grew (this reversable growth-evaporation dynamics could be clearly seen in the *in situ* TEM experiments, by setting the substrate temperature $T$ below and above the critical deposition temperature $T_o$). The $\Delta\gamma_{LR}$ value thus obtained is:

$$\Delta\gamma_{LR} = -\frac{\tanh^{-1}(EE)\rho^2 RT \Delta H_{sub}^2 \Delta T^2}{2\pi\gamma_g^2 N_A M^2 T_0^2} =  \qquad \text{(Eq. S19)}$$

$$-\frac{\tanh^{-1}(0.73)\ 6.24^2\left[\frac{g}{cm^3}\right]^2 10^{48}\left[\frac{cm}{\text{Å}}\right]^6 8.314\left[\frac{J}{mol\,K}\right]\ 6.24\cdot10^{18}\left[\frac{eV}{J}\right]\ 553[K]\ 156.6^2\left[\frac{kJ}{mol}\right]^2 10^6\left[\frac{J}{kJ}\right]^2 3.89\cdot10^{37}\left[\frac{eV}{J}\right]^2\ 44^2[K]^2}{2\cdot\pi\cdot186^2\left[\frac{mN}{m}\right]^2 10^{-6}\left[\frac{N}{mN}\right]^2\ 3.89\cdot10^{37}\left[\frac{eV}{J}\right]^2\ 10^{40}\left[\frac{m}{\text{Å}}\right]^4\ 6.022\cdot10^{23}\left[\frac{1}{mol}\right]\ 52^2\left[\frac{g}{mol}\right]^2 509^2\ [K]^2} =$$

$$= -4\cdot10^{-3}\left[\frac{eV}{\text{Å}^2}\right]$$

This is slightly larger (in absolute value) than the estimated DFT value shown above (Section S1.12) in Eq. S18, $-1.7\cdot10^{-3}\frac{eV}{\text{Å}^2}$, but at a remarkably similar order of magnitude considering the variations in many of the parameters. The values in Table S3, which were calculated by DFT in a different way (section S1.11) are one order of magnitude smaller. These discrepancies could be attributed to a large uncertainty in the experimental $T_0$-$T$ difference, and large variation in the evaporation Te enthalpy and gas-solid interface energy values found in literature. Notably, the sign of $\Delta\gamma_{LR}$ matches the expected trend of favorable nucleation of left-handed Te NWs on the (001) basal plane of ReSe$_2$. The plot in Fig. 4B uses the experimental value of $\Delta\gamma_{LR}$ to fit the *EE* obtained at the experimental temperatures and extrapolates the expected values of *EE* for the entire range of sample temperatures following Eq. S13.

### S2 Supplementary Analysis

Plane-view sample analysis

To determine the handedness along the NWs without the need of cutting a FIB lamella, a ReSe$_2$ flake was transferred onto a holey membrane chip as described in S1.7. Te NWs were grown on the suspended ReSe$_2$ flake by CVD, resulting in a plane-view sample. The sample can be placed inside the TEM with the Te NWs facing either the beam or the detector on the opposite side (fig. S9 A and B). For each sample placement, the ReSe$_2$ basal plane orientation was assigned from the diffraction image (fig. S9E), based on a previous reported method (*64*), using the Re$_4$ parallelogram orientation as seen in the HAADF-STEM image in fig. S9F. To check whether the handedness of each NW remains the same along its growth axis, Te NWs were grown atop a ReSe$_2$ flake suspended over a holey SiN TEM grid. Using the tilt-series method (this time tilting the NWs about their axis by 30°, as in (*38*)), the handedness of the NWs was analyzed along their entire length, and was observed to remain constant along all of them. An example of a right-handed Te NW grown on the ReSe$_2$ (00$\bar{1}$) basal plane, exhibiting constant handedness, is presented in fig. S10.

The handedness determination in the plan-view orientation was based on the Te projections in fig. S14. In some cases, the Te NWs were so thin that the imaging of the NWs grown on the ReSe$_2$ resulted in Moiré pattens arising from the superposition of the Te and ReSe$_2$ lattices, and then the atomic ordering of the Te NWs could not be resolved. Therefore, the handedness determination was assigned for less than 10 NWS, the calculated *EE* value was 43% (fig. S11). The handedness analysis of the NWs is presented in fig. S12. Interestingly, the bottom panel of the image shows that there is a crack in the ReSe$_2$ flake, suggesting that Te can grow on both sides of the suspended ReSe$_2$ flake. When tilting the Te NW that grew on the edge of the flake, a clear grain boundary with the morphology of the cracked ReSe$_2$ flake is observed, each of the Te grains having an opposite handedness. To further investigate this, two Te NWs were analyzed, one growing on the (001) basal plane of ReSe$_2$ and the other on the (00$\bar{1}$) basal plane (fig. S13). The Te NW growing on the (001) plane is left-handed (L-Te) while the NW growing on the opposite side of the ReSe$_2$ is right-handed (R-Te). The two options for the orientation of the sample are presented in fig. S9. To establish whether the Te NW lay above or below the flake, we varied the focal plane continuously and observed which fringes appeared closer to or farther away from the detector.

Grain boundaries as observed by *ex situ* and *in situ* TEM experiments

NWs can form over the entire surface of the ReSe$_2$ flake, and adjacent NWs can grow into one another and merge together when nucleating in proximity. If the merging NWs have different handedness or orientation, a distinct grain boundary should form, as the triangular helical chains cannot create a close-packed hexagonal structure. These grain boundaries are observable in both cross-sectional and planar views (fig. S12). *In situ* TEM and SEM movies support having NWs that merge together during growth without fully coalescing into a single crystal. The recorded movie S7 shows that when NWs grow in opposite directions and bump into each other, they sometimes form a grain boundary. This grain boundary can change its position over time without disappearing (fig. S21). The occurrence of these non-mixing NWs can be interpreted as the contact between Te NWs having different handedness or orientation. This means that if a NW were to change its handedness during growth, a grain boundary should have been observed between the regions having different handedness. As mentioned earlier in the plane-view analysis, 10 NWs showed no such grain boundaries along their growth axis, supporting our suggestion that once the initial nucleus is formed and starts to elongate, the crystal maintains its original crystallographic structure and will not change its handedness as it keeps growing. In some cases, as observed in the *in situ* SEM movies (movie S3, S4) nanowires can expand laterally until they touch one another. When preparing the cross-sectional lamellae, some of the wider nanowires may therefore represent two adjacent nanowires that have merged laterally. Given the short reaction time and low substrate temperature, these merged regions are unlikely to anneal into single crystal because that would require an entire grain to flip its handedness, which would have an insurmountable activation energy. Thus, the existence of a grain boundary can indicate that these are separate nanowires that merged together. Once a grain boundary was identified, each side of the boundary was treated as a separate grain, and the handedness of each grain was determined independently. Thus, the fact that a change of handedness was not observed within individual NWs further supports that each Te NWs is indeed a single crystal, without changing its handedness along its growth axis.

Interfacial strain between the Te NWs and the ReSe$_2$ substrate

To calculate the interfacial strain between the Te NWs and the ReSe$_2$ in both the cross-sectional and longitudinal lamella Eq. 20 was used, where the $d_{sub}$ represents the d-space of the substrate (ReSe$_2$) and $d_{epi}$ represents the d-space of the epilayer (tellurium). The d-spaces values where taken from the cif files of each material.

$$f(mismatch) = 100\% \text{ x} \frac{d_{sub} - d_{epi}}{d_{sub}} \qquad \text{(Eq. S20)}$$

The strain within each crystal (ReSe$_2$ or tellurium) was determined by Eq. S21. Where the $d_{measured}$ was obtained from the FFT analysis of the STEM images of the Te/ReSe$_2$ heterostructure and the $d_{theory}$ was taken from the cif files of each material.

$$strain = 100\% \text{ x} \frac{d_{measured} - d_{theory}}{d_{measured}} \qquad \text{(Eq. S21)}$$

For the cross-sectional lamella, the epitaxial relations are approximately $(\bar{4}21)_{ReSe_2} || (2\bar{1}\bar{1}0)_{Te}$. The ratio between the number of ReSe$_2$ and Te d-spacing is 4:3, therefore the coincidental mismatch is:

$$f(mismatch) = 100\% \text{ x} \frac{4 \cdot 1.6446 - 3 \cdot 2.229}{4 \cdot 1.6446} = 1.6\% \qquad \text{(Eq. S22)}$$

Based on the d-spaces measured in the FFT image of each crystal the measured strain within the lateral planes of ReSe$_2$ is -1.92% and of Te is -1.1%.

For the longitudinal lamella, the epitaxial relations are $(001)_{ReSe_2} || (0001)_{Te}$, the ratio between the number of ReSe$_2$ and Te d-spacing is 1:1, therefore the mismatch is:

$$f(mismatch) = 100\% \text{ x} \frac{5.4873 - 5.926}{5.926} = 8.04\% \qquad \text{(Eq. S22)}$$

Based on the d-spaces measured in the FFT image of each crystal the measured strain within the lateral planes of ReSe$_2$ is 4.7% and of Te is 3.2%.

The detailed epitaxial relations in the Te/ReSe$_2$ cross-sectional interface

In terms of symmetry, the Te/ReSe$_2$ system has some complexity to be considered. As mentioned, the Te crystal can have two possible orientations with respect to the ReSe$_2$ basal plane, irrespective of their handedness. Although Te has three $C_2$ rotation symmetry axes perpendicular to chain axis x, the ReSe$_2$ surface itself does not have any rotation or mirror symmetry. Therefore, each enantiomer L-Te or R-Te can have two different orientations with respect to the ReSe$_2$ substrate (depending on whether the Te chains are along the [0001] or [000$\bar{1}$] c-axis of the NW). This generates overall eight different Te-ReSe$_2$ interface combinations, where Te can be L or R, with two different orientations of the c-axis with respect to the ReSe$_2$, and the ReSe$_2$ surface itself can be either (001) or (00$\bar{1}$). A summary of all the epitaxial relations and their frequency for a random set of samples that were comprehensively analyzed is shown in Tables S1 and S2, respectively. We found that the Te NWs orientation (grown along the [0001] or [000$\bar{1}$]), irrespective of their handedness, also shows some degree of selectivity, but not as significant as the enantiomeric excess of chiral epitaxy. On the other hand, Te NWs with a specific orientation can show higher enantioselectivity, with *EE* values up to 93%.

<u>Molecular dynamics simulations analysis of the possibility of handedness flip in small Te embryos</u>

To assess the kinetic stability of the Te-ReSe$_2$ heterostructure and determine whether a handedness flip can occur at initial stages of growth, we performed *ab initio* molecular dynamics simulations (see details in S1.11). When simulating a right-handed Te embryo on ReSe$_2$ (001) (the less stable configuration on this surface), consisting of 72 Te atoms, it undergoes a spontaneous transition to the more stable left-handed configuration within 2.5 ps. Conversely when starting with left-handed Te few-atom embryos on ReSe$_2$ (001), no such transition occurs (Fig. 4D).

Interestingly, when the cluster size was increased to 144 Te atoms (arranged into two layers of chains), no spontaneous transition occurred within the simulation timeframe, even for the less stable enantiomer. These findings suggest that, while very small embryos can interconvert to the thermodynamically preferred handedness, larger clusters become kinetically trapped in the thermodynamically preferred handedness. This supports a mechanism where enantioselectivity is established at the earliest stages of nucleation. While our analytical model (Eq. 1) assumes a pre-equilibrium between embryos, these simulation results provide a microscopic basis for knowing how the system reaches the preferred enantiomeric state before the growth chirality becomes irreversible.

<u>The conditions for achieving chiral epitaxy in other synthetic systems</u>

For any candidate system, three conditions must be satisfied: the substrate surface must lack mirror symmetry, the epilayer must be intrinsically chiral in its bulk structure, and the interface energy difference $\Delta\gamma_{LR}$ between enantiomers must be large enough to produce significant selectivity. The substrate surface should be enantiomerically pure enough, namely without too many defects, especially chiral inversion domains (*64*). Computational screening of interface energies could help identify combinations with large $\Delta\gamma_{LR}$ before undertaking growth experiments. Furthermore, in materials where chirality does not readily change during growth, the entire templating substrate surface does not need to be chiral to achieve chiral epitaxy, only the nucleation sites. This opens potential for spatially patterned chiral nucleation sites on otherwise achiral surfaces to enable integration of chiral elements into conventional device fabrication.

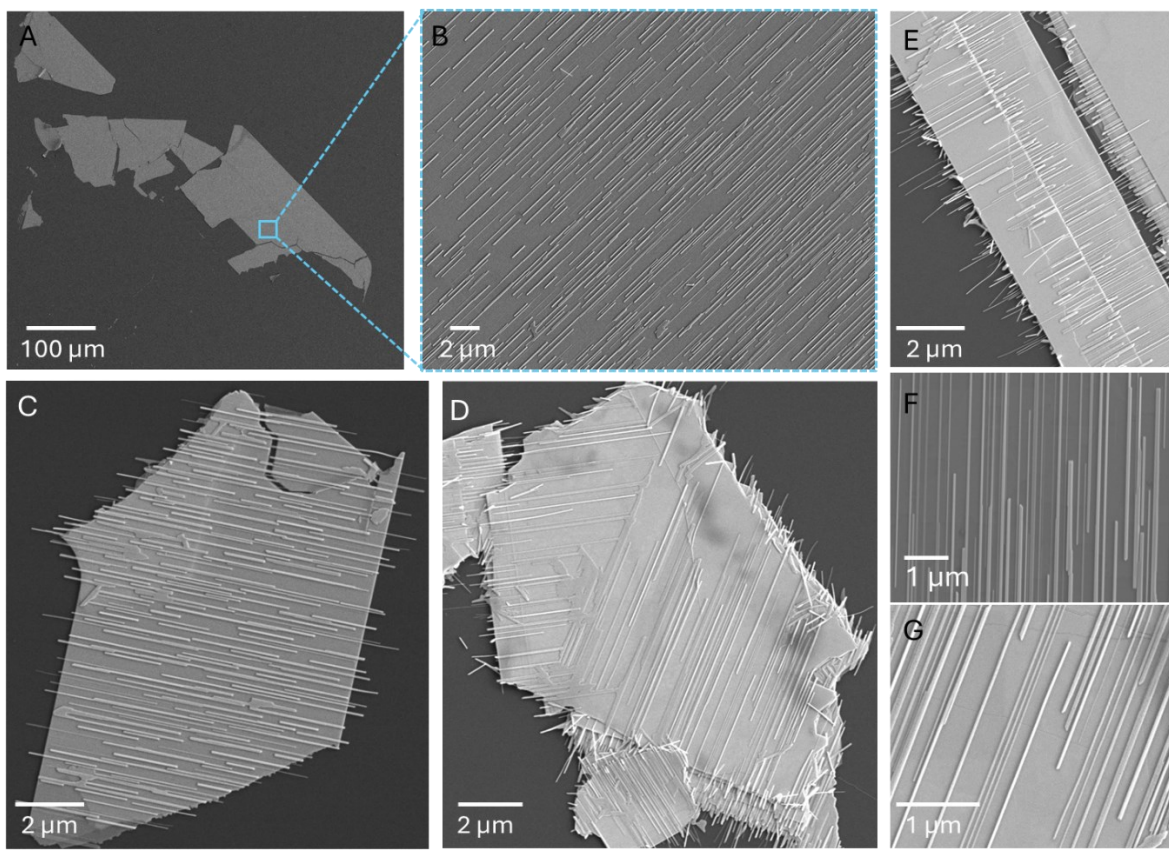

**Fig. S1. Te NWs grown on ReSe₂ as observed in SEM images.** (**A**) Low-magnification of large ReSe₂ flakes, with Te NWs growing on them. (**B**) High-magnification of the aligned Te NWs grown on the marked blue rectangle in (A). (**C** to **G**) Aligned Te NWs growing on various ReSe₂ flakes, in various magnifications. (**C**) Te NWs growing over a crack in the ReSe₂ flake, which was formed during the exfoliation. (**D**) NWs growing on a ReSe₂ flake in two distinct directions, indicating the presence of a domain boundary in the flake, as indicated in a previous study (*65*). (**E**) Te NWs growing on a step in the middle of the ReSe₂ flake and on the edges of the flake. (**F**, **G**) High-magnification images of aligned fine tellurium nanowires grown on ReSe₂.

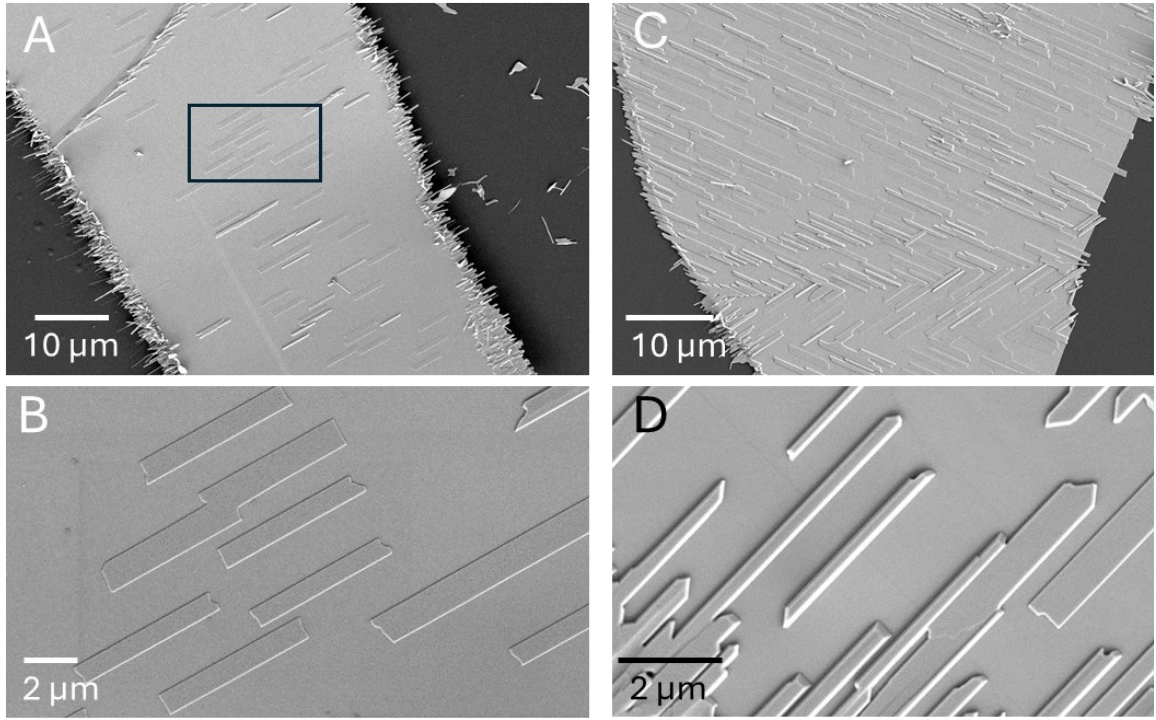

**Fig. S2. Different morphologies of wide Te NWs (nanoribbons or nanobelts) growing on ReSe₂ as observed in SEM images.** (**A**) Low-magnification of Te nanobelts grown on a ReSe₂. (**B**) High-magnification of the marked rectangle in A, showing the detailed morphology of the nanoribbons, having a pointed edge tip. (**C**) Te nanoribbons growing on a ReSe₂ flake with a domain boundary, where two growth orientations form a fishbone-like structure at the bottom of image. (**D**) High-magnification of Te nanobelts with unique asymmetrical edges.

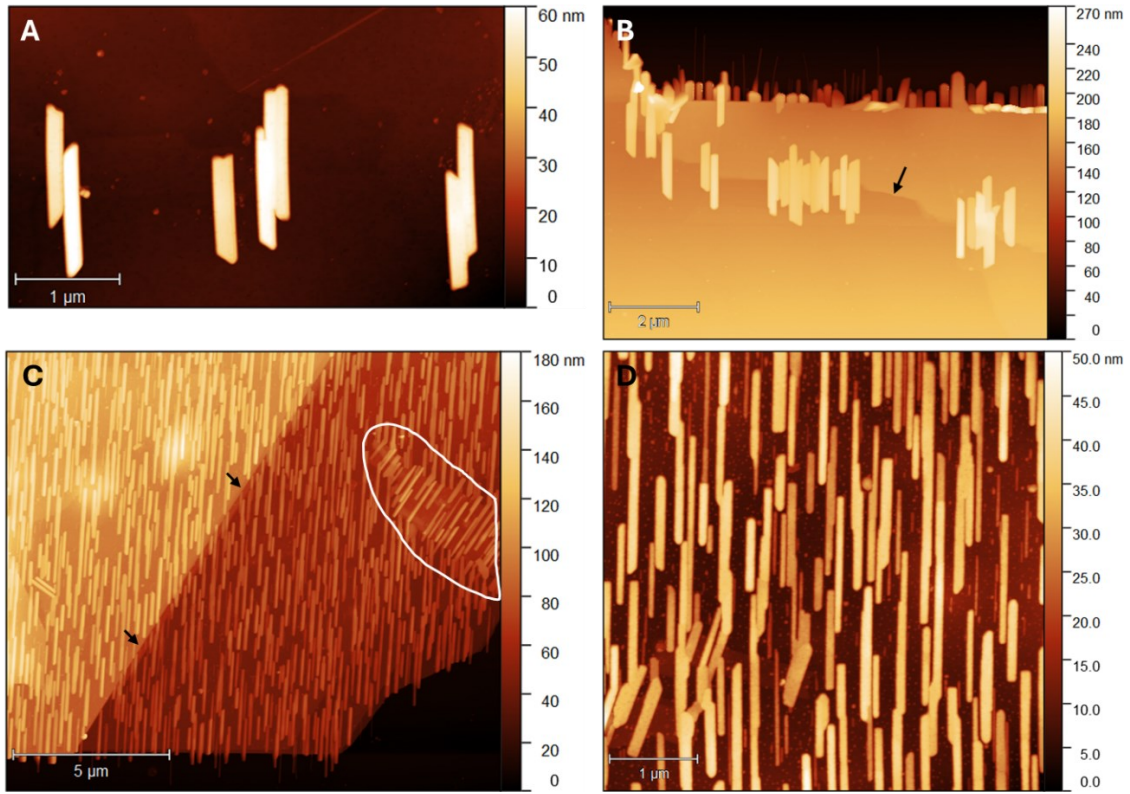

**Fig. S3. AFM images of Te NWs grown on ReSe₂.** (**A**) Te NWs growing in the middle of a ReSe₂ flake with a height of 40-60 nm. (**B**) Te NWs growing on the edge of the ReSe₂ flake and step generated during exfoliation, indicated by the black arrow. Height of the NWs ranges between 40-75 nm. (**C** and **D**) Te NWs grown on ReSe₂, where there is high coverage of Te NWs all over the flake. Height of the NWs varies between 10 and 40 nm. In image (**C**) there is a clear step in the middle of the flake, marked by two black arrows. Marked in white is a smaller flake that adhered to the larger ReSe₂ flake. The NWs have a different growth orientation, because the adhered flake has a different orientation from the underlying flake.

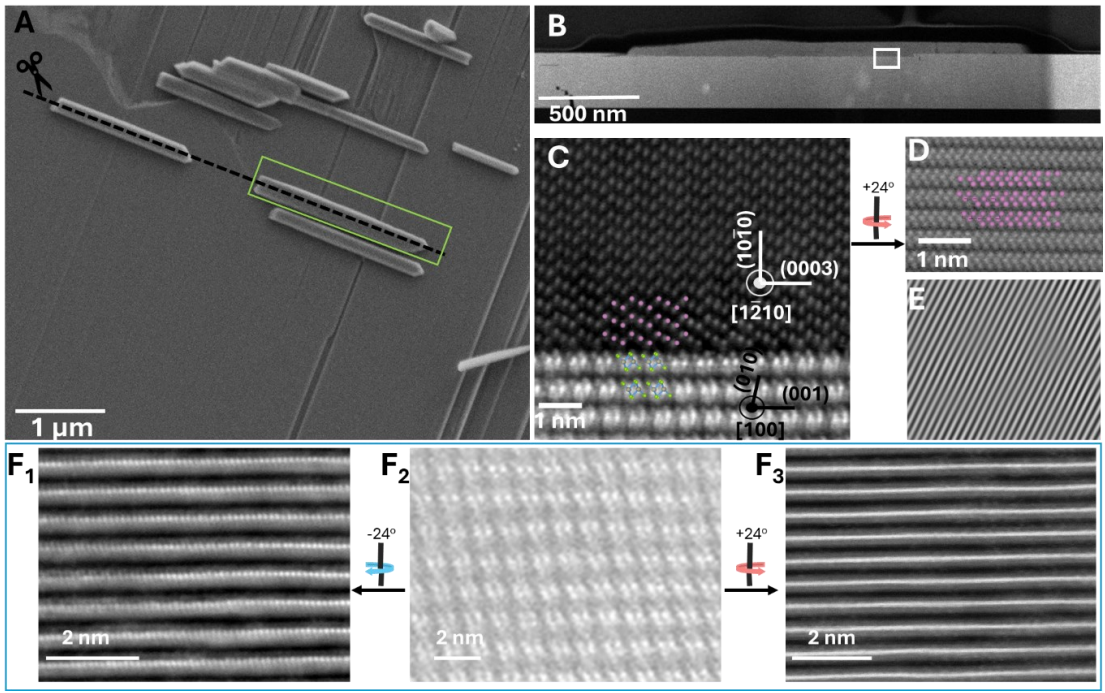

**Fig. S4. Handedness of Te NWs along the NW.** (**A**) SEM image of the two NWs that were cut into a longitudinal lamella. (**B**) Low magnification of the right NW in image (**A**). The area from which images (**C and D**) were acquired is marked in white. (**C**) HAADF-STEM image of the longitudinal epitaxial relations between the Te and ReSe₂, where the Te bottom atoms are situated above the Re atoms. (**D**) HAADF-STEM image of the Te atoms' arrangement after 24º tilt, showing left-handed Te, as in the model in fig. S5. (**E**) The Inverse Fast Fourier Transform of image (**C**), showing the commensuration of the interface, with 1:1 epitaxial relation with no dislocations. (**F**) HAADF-STEM images of the tilting to ±24º of the ReSe₂ (001) basal plane, based on the model in fig. S6. (**F₂**) The orientation of the ReSe₂ before tilting and after tilting the lamella to the left (**F₁**) where the Re atoms are clearly observed and tilting to the right (**F₃**), where the Re atoms are overlapping and cannot be distinguished.

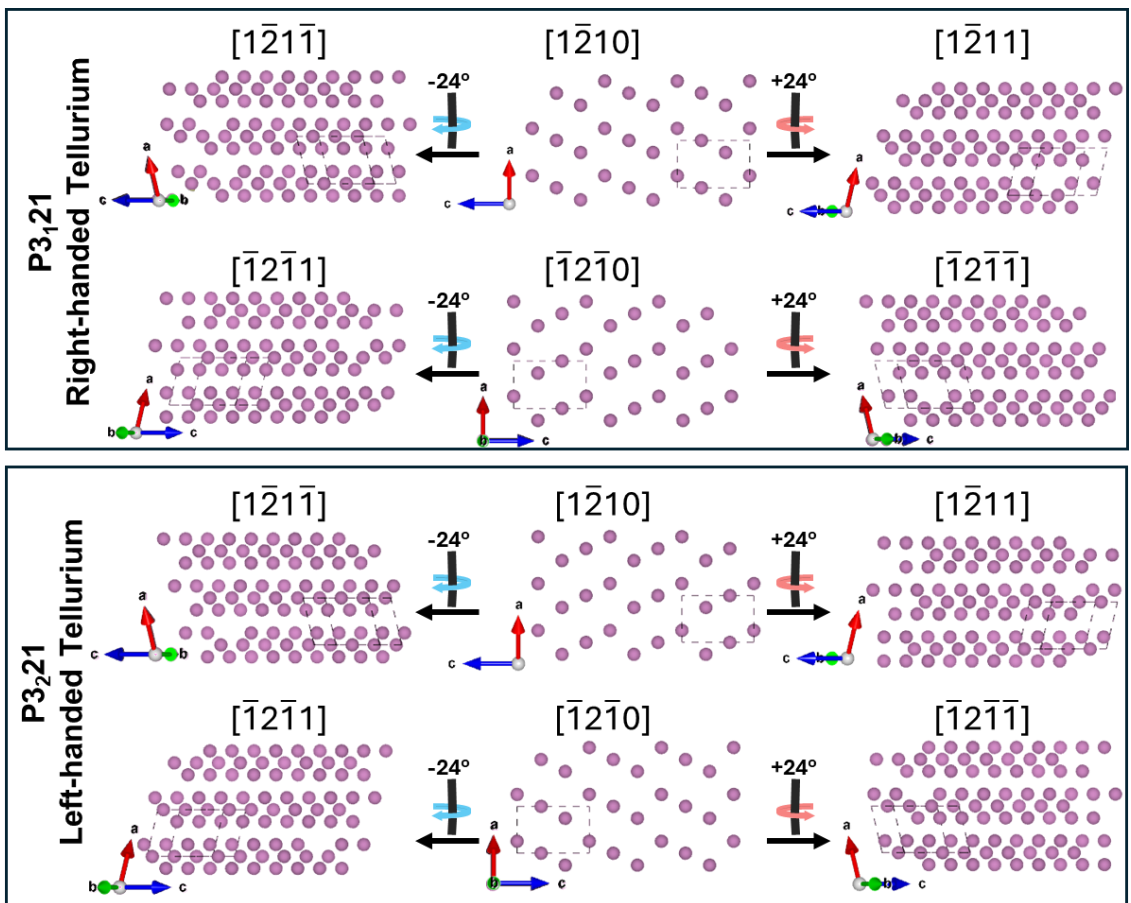

**Fig. S5. Tilt series projections of right- and left-handed Te for a lamella along the NWs.** Top (bottom) box shows the tilting series for right-handed (left-handed) longitudinal view of Te NWs, used to determine the Te handedness in fig. S4.

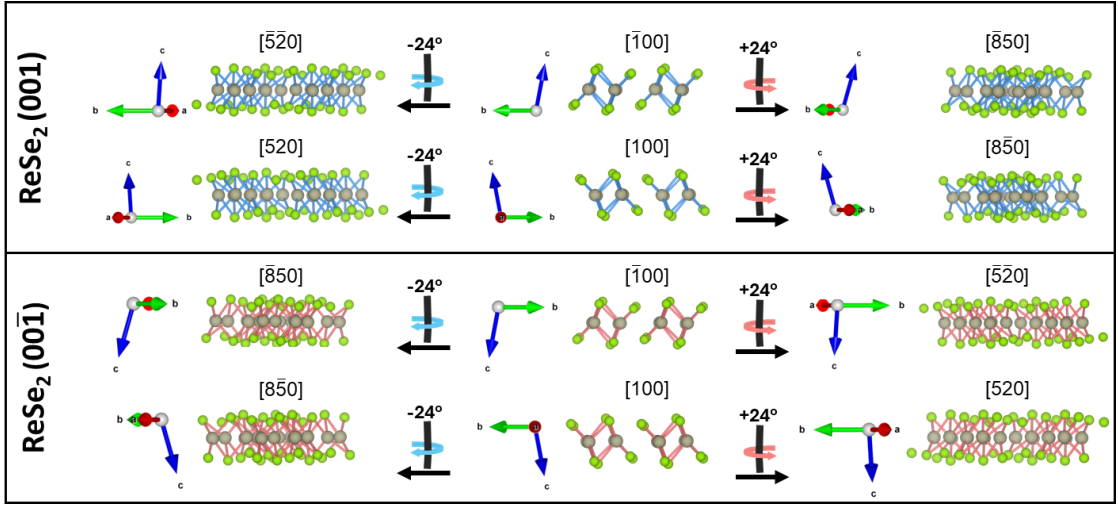

**Fig. S6. Tilt series projections of ReSe₂ (001) and (00$\overline{1}$) for a lamella cut along the Te NWs.** Top (bottom) box shows the tilt series for identifying ReSe₂ (001) basal plane (ReSe₂ (00$\overline{1}$) basal plane) when viewing the longitudinal cross-section of the Te NWs, used in fig. S4.

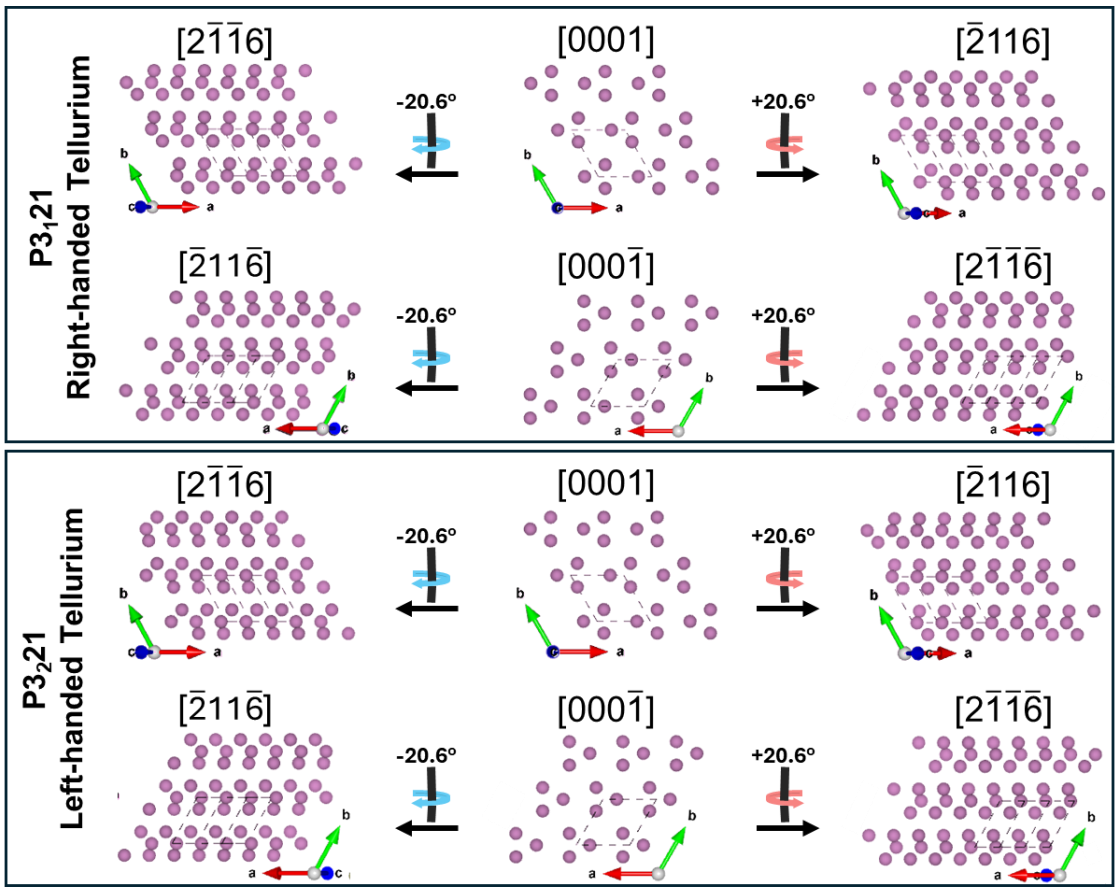

**Fig. S7. Tilt series projections of right- and left-handed Te for a lamella across the NWs.** Top (bottom) box shows the tilting series for right-handed (left-handed) Te NW cross sections, used to determine the Te handedness as in Fig. 2.

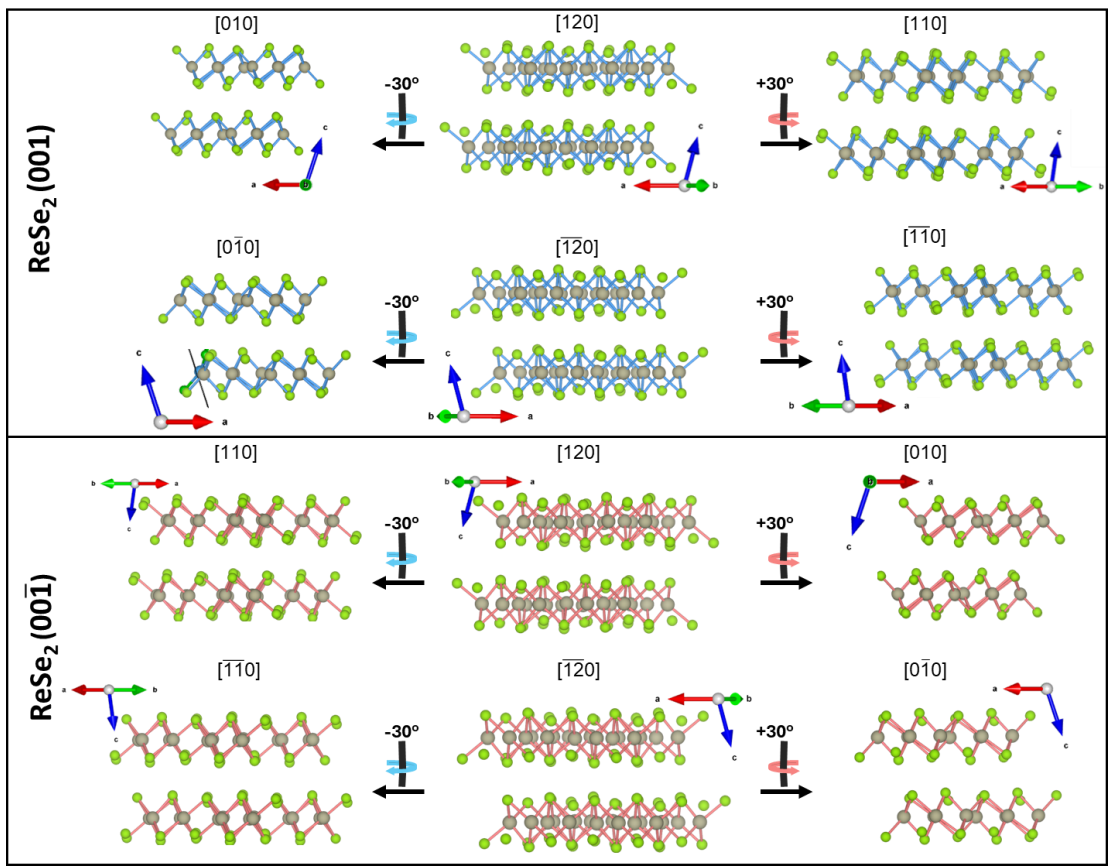

**Fig. S8. Tilt series projections of ReSe₂ (001) and (00$\bar{1}$) for a lamella cut across the Te NWs.** Top (bottom) panel shows the tilt series for ReSe₂ (001) basal plane (ReSe₂ (00$\bar{1}$) basal plane), when viewing the cross-section of Te NWs, as used in Fig. 2.

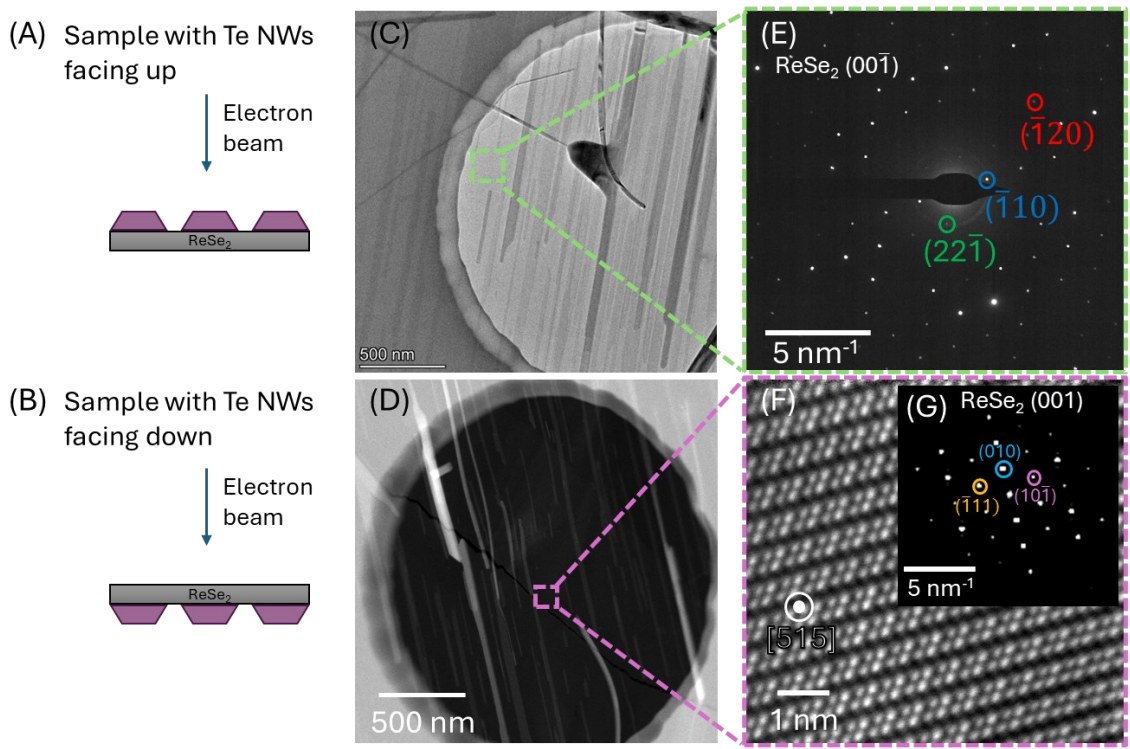

**Fig. S9. Plan-view orientation of Te NWs grown on a ReSe₂ membrane.** (**A**) and (**B**) shows the sample orientation relative to the electron beam in the top and bottom panels respectively. In (**A**) the NWs are observed on top of the ReSe₂ membrane, while in (**B**), the NWs are observed below the ReSe₂ membrane. (**B** and **D**) Low-magnification of a plan-view sample, where Te NWs were grown on a suspended ReSe₂ flake. (**E**) Electron diffraction pattern acquired from the green-marked square in (**B**). The low symmetry of the ReSe₂ allows determination of its basal plane orientation by assigning the planes in the diffraction, in this case, it has the $(00\bar{1})$ basal plane orientation. (**F**) High-resolution HAADF-STEM image of the atomic Re chains orientation, with the zone axis of [515]. The inset (**G**) shows the fast Fourier transform (FFT) analysis of image (**F**), confirming that the ReSe₂ orientation corresponds to the (001) basal plane in this viewing in this direction.

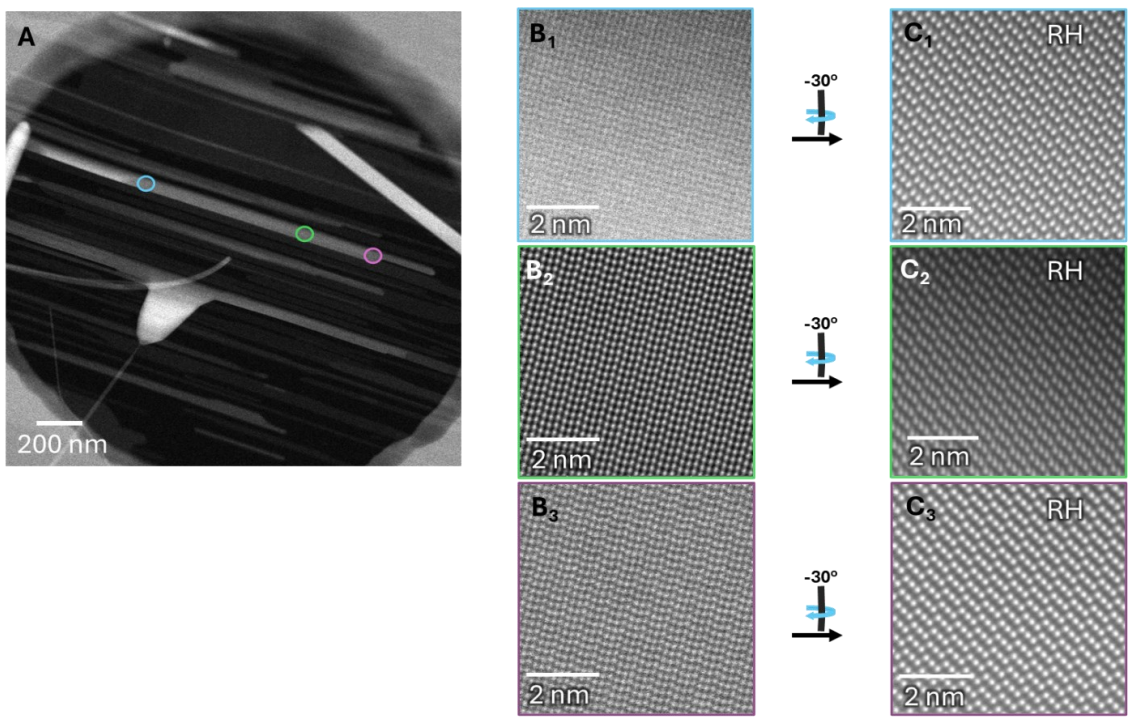

**Fig. S10. Handedness conservation along the NW's length.** (**A**) The low-magnification image of a Te NW on top of ReSe$_2$ with (00$\bar{1}$) basal plane orientation. Three positions along the nanowire, marked in light blue, green, and purple in (A), were analyzed to assess the atomic orientation. (**B**) Shows the orientation of Te atoms at these marked positions before tilting, with images B$_1$ to B$_3$ confirming the same atomic arrangement at all three points. (**C**) Displays the corresponding atomic orientations after tilting the sample 30° to the left. Images C$_1$ to C$_3$ demonstrate a consistent Te atom arrangement, all having right handedness.

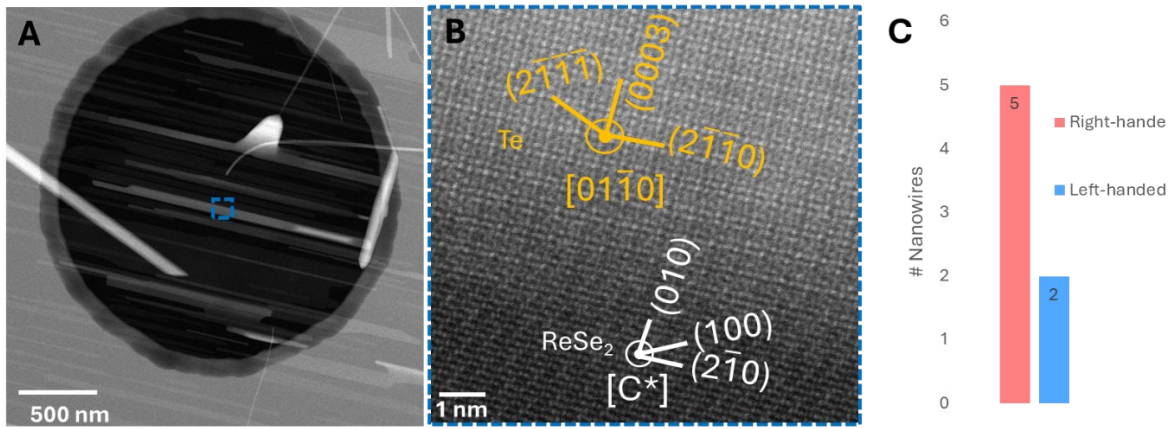

**Fig. S11. Enantiomeric excess of plan-view Te NWs.** (**A**) Low magnification of Te NWs grown on ReSe$_2$ suspended membrane. The sample is placed with the ReSe$_2$ facing the electron beam. (**B**) High resolution HAADF_STEM image of the marked blue square in image (**A**) showing the epitaxial relations between the Te and ReSe$_2$ from plan-view. The two crystals generate a Moiré patten visible where the Te NWs are thin enough, making it difficult to assign the NWs handedness. (**C**) The recorded enantiomeric excess obtained for thick enough Te nanowires (where no Moiré pattern is visible) within the plane-view sample, showing an excess of right-handed NWs over left-handed NWs as expected.

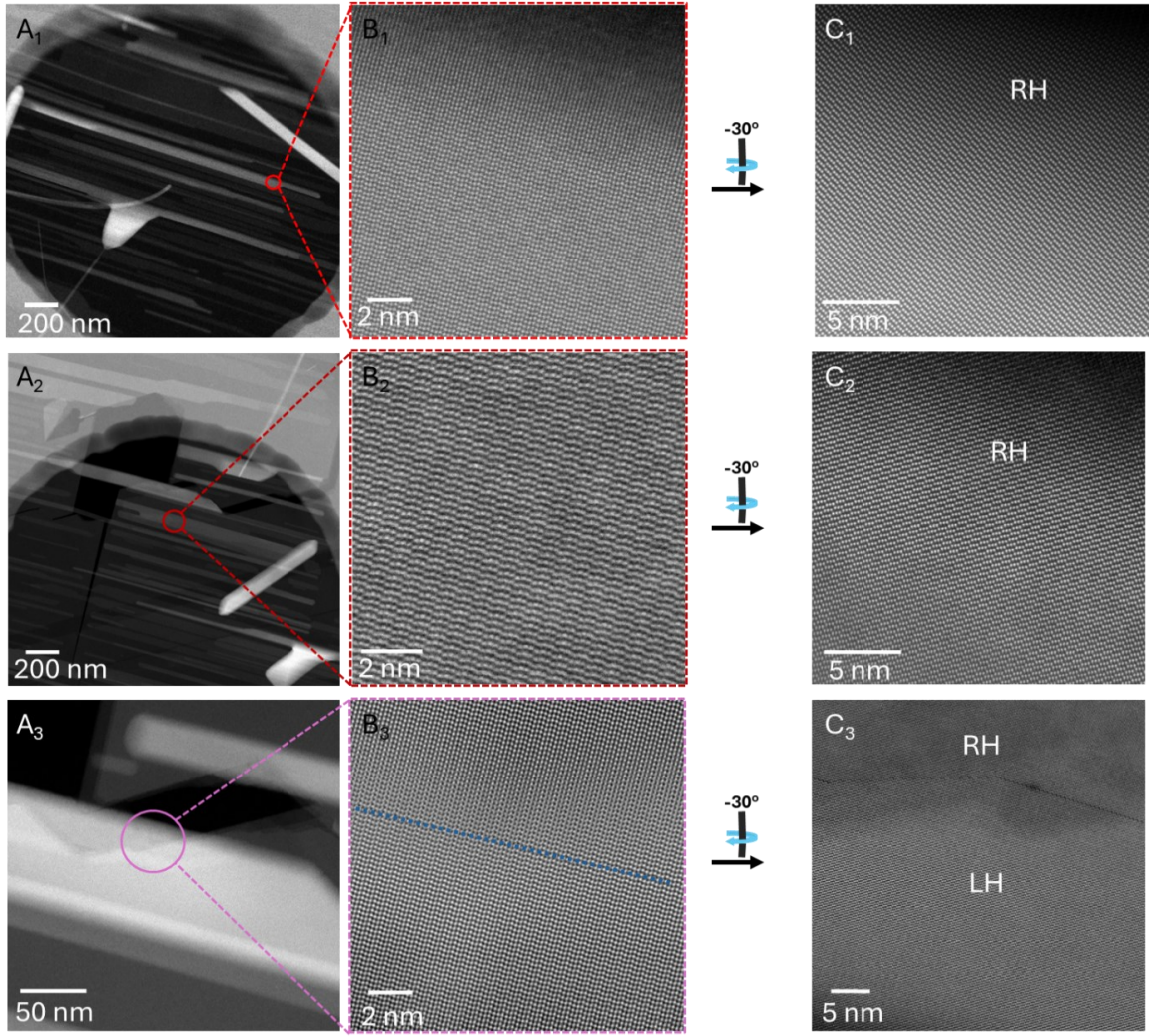

**Fig. S12. Determining the handedness of plane-view Te NWs. (A to C)** HAADF-STEM images of the handedness determination of plan-view Te NWs grown on a $ReSe_2$ suspended flake. ($A_1$ to $A_3$) Low magnification images of the analyzed Te NWs. In $A_3$, a clear crack in the $ReSe_2$ flake is presented. ($B_1$ to $B_3$) High-resolution images of the Te atoms arrangement. In $B_3$ there is a grain boundary in the middle, where the Te atoms switch to the other orientation, marked with a dashed blue line. ($C_1$ to $C_3$) High-resolution HAADF-STEM images of the Te NW after tilting by 30º to the left, handedness was determined based on fig. S14. $C_1$ and $C_2$ show right-handed orientation, while in $C_3$ there is a grain boundary, separating the top right-handed and the bottom left-handed NWs. The grain boundary morphology is aligned with the edge morphology of the cracked $ReSe_2$.

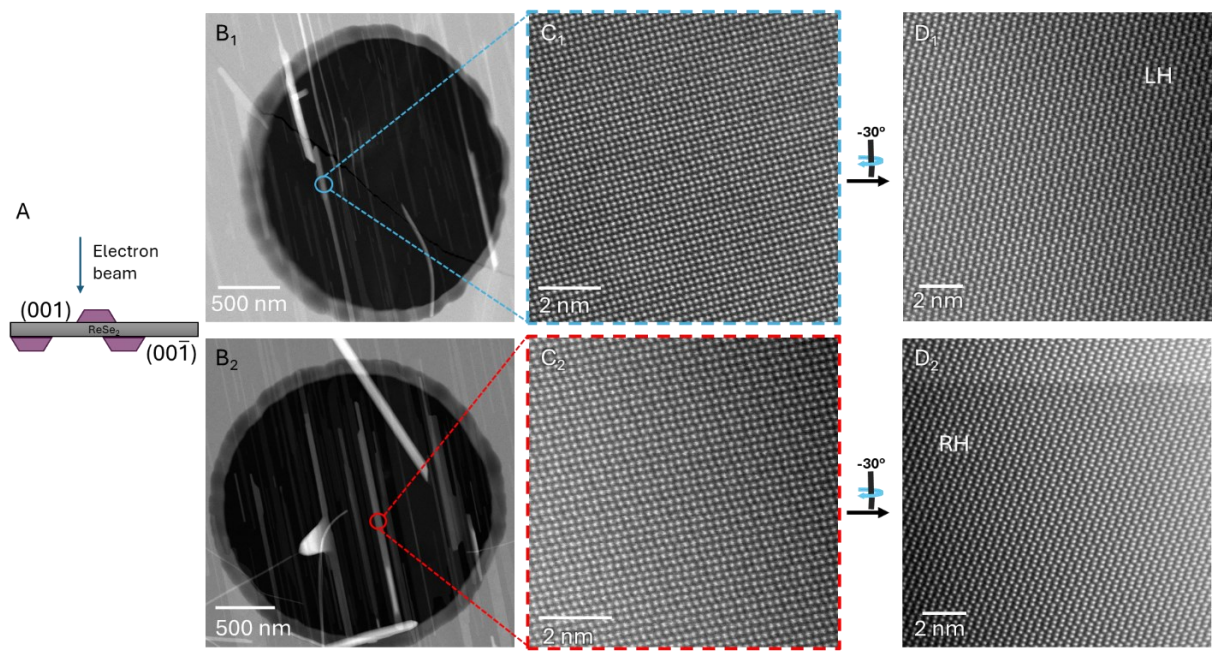

**Fig. S13. NWs growing on opposite sides of the ReSe₂ membrane exhibit opposite handedness.** (**A**) Schematic showing the sample orientation relative to the electron beam; the position of each NW, closer or farther from the beam, was determined based on focus distance. (**B to D**) HAADF-STEM images of the handedness determination of plane-view Te NWs grown above and below a ReSe₂ suspended flake. (B₁ to B₂) Low magnification images of the measured NWs marked with a circle, in B₁ the NW grew above the ReSe₂ while in B₂ the NW grew below the ReSe₂. The ReSe₂ orientation relative to the electron beam is determined based on Fig.S9. (C₁ and C₂) High-resolution images of the Te atoms' arrangement. (D₁ and D₂) Images of the Te NWs after tilting them by 30° to the left. The NW in D₁ is left-handed, while the NW in D₂ is right-handed, determined based on the model in fig. S14. Both NWs grew on the expected ReSe₂ basal plane, with the left-handed NW growing on (001) ReSe₂.

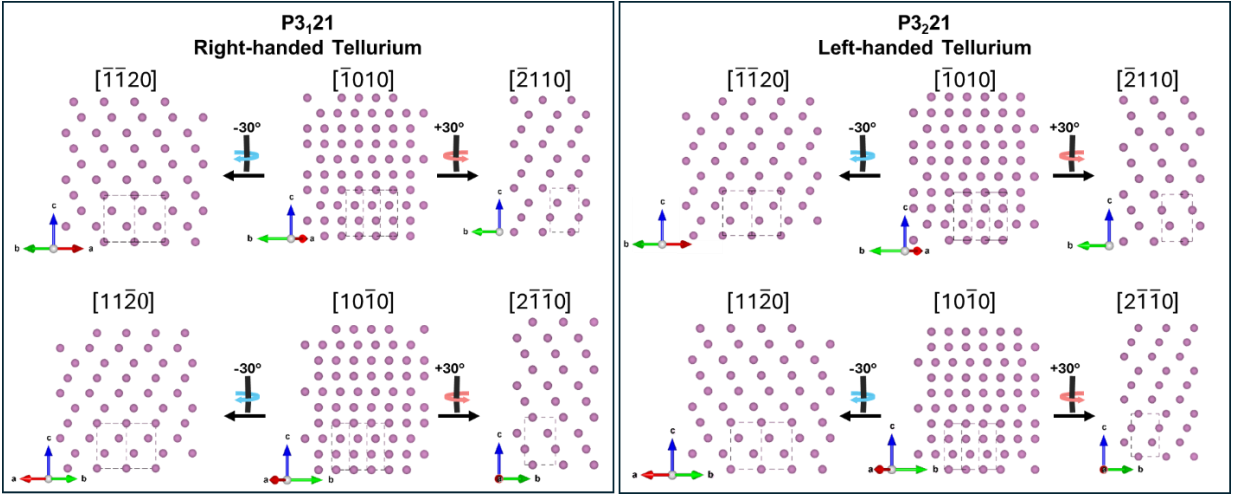

**Fig. S14. Tilt series projections of right- and left-handed Te for a top view of the NWs.** Left (right) box shows the tilting series for right-handed (left-handed) plane-view orientation of Te NWs, used to determine the Te handedness in figs. S9-S13.

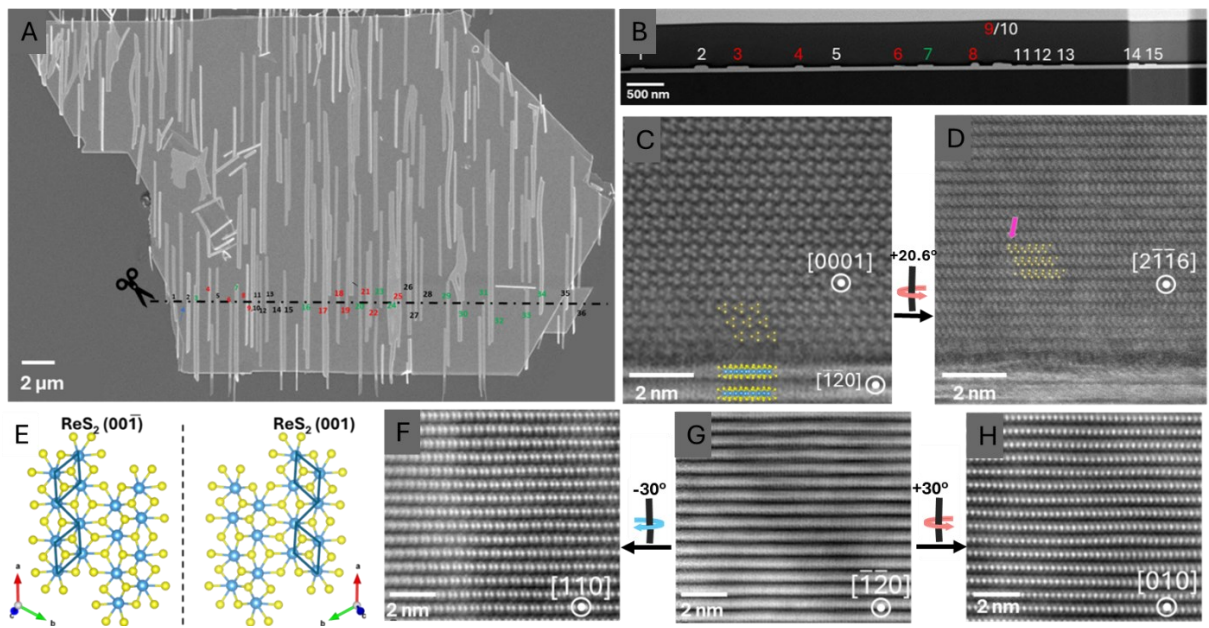

**Fig. S15. Te NWs grown on ReS₂ and characterization of their epitaxial relations**: (**A**) SEM image of the Te NWs grown on ReS₂. The green NWs are left-handed, and the red ones are right-handed. (**B**) Low magnification of the cross-section lamella showing the interface between the Te NWs and the ReS₂. At the interface the Te atoms are smudged, most probably due to damage during lamella fabrication; (**C**) High resolution HAADF-STEM image of the cross-section of Te NW on ReS₂ substrate. The atomistic models of Te and ReS₂ are overlaid on the image (**D**) HAADF-STEM tilted image of NW number three, showing a right-handed NW. (**E**) Atomistic models of the (001) and (00$\bar{1}$) ReS₂ planes, the blue and yellow atoms represent the Re and S atoms, respectively. (**F** to **G**) HAADF-STEM images of the ReS₂ with (00$\bar{1}$) basal plane. Image (G) shows the untilted projection, while (F) and (H) correspond to projections tilted 30° to the left and right, respectively.

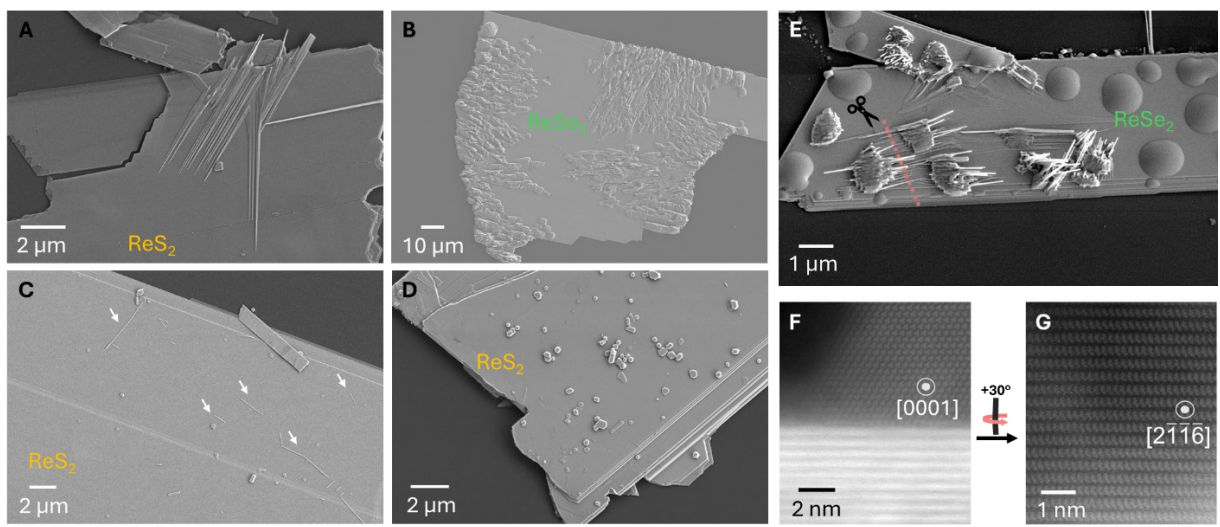

**Fig. S16. Se NWs and nanocrystals grown on ReS₂ and ReSe₂.** (**A, C**) Se NWs grown on ReS₂, in (**C**), the fine NWs are indicated by white arrows. (**B**) Se crystals growing in various orientations on ReSe₂ flake. (**D**) Se hexagonal nanocrystals grown on ReS₂. (**E**) Se NWs growing out of Se droplets on ReSe₂. The pink dotted line indicates the location where the lamella was cut. (**F**) Cross-section view of the Se NW on ReSe₂, the zone axis of the ReSe₂ is different from the one observed for Te. (**G**) Projection tilted by 20.6° of the Se NW in (**F**), showing a left-handed NW.

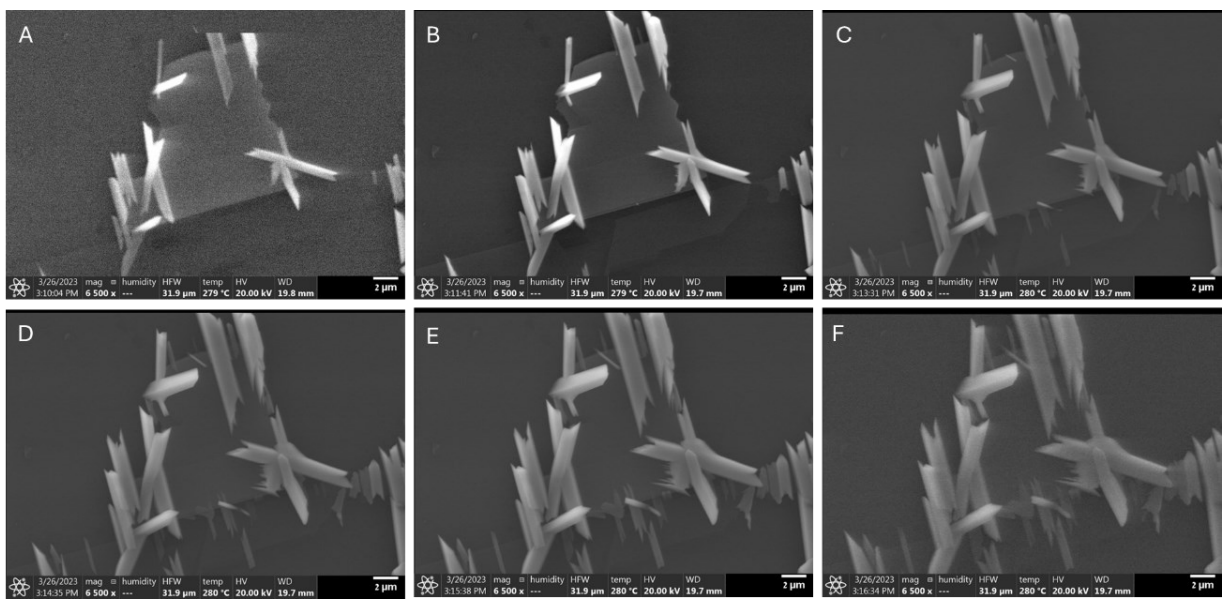

**Fig. S17. Fast heating rate leads to formation of hollow Te nano- and microtubes. (A** to **F)** SEM image sequence of Te micro- and nano-tubes growth inside environmental SEM, with high heating rate of the crucible source. The image sequence corresponds to movie S1.

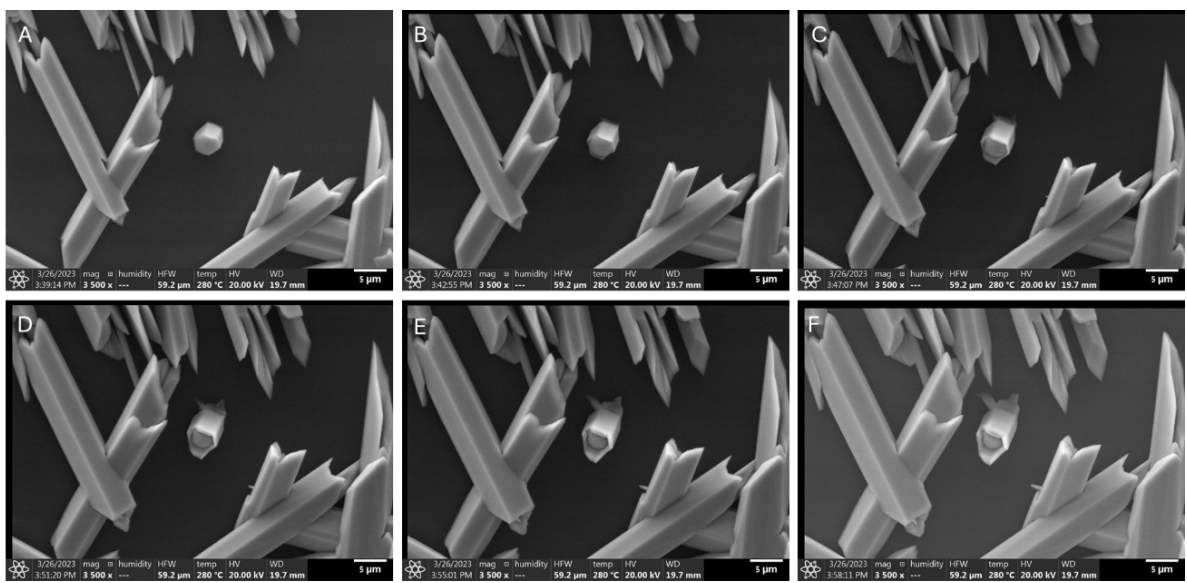

Fig. S18. SEM image sequence of a Te microtube growing out of plane, showing the tube growth mechanism. The image sequence corresponds to movie S2.

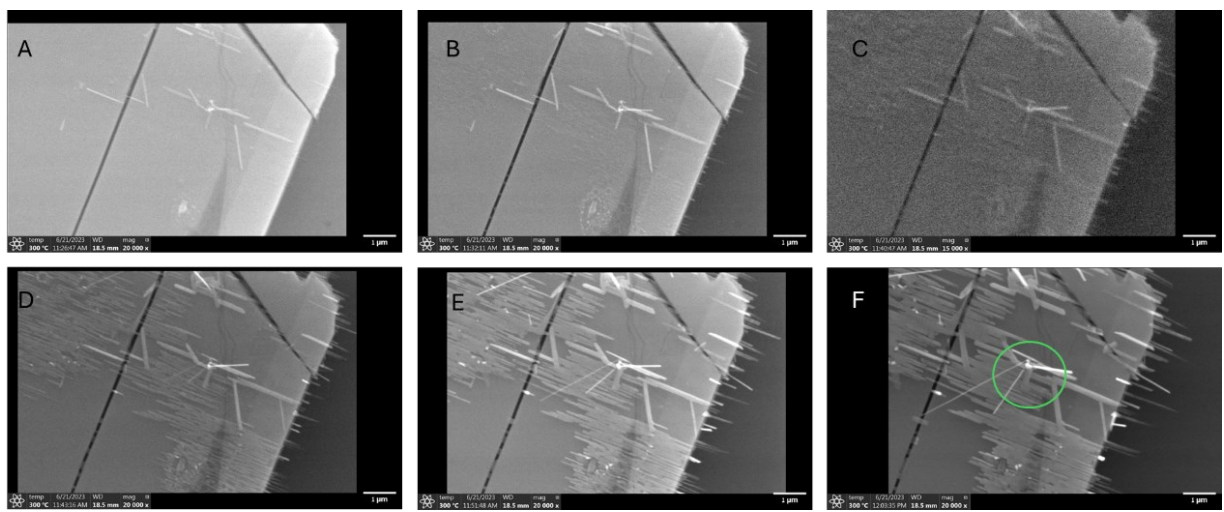

**Fig. S19. Real-time growth of Te NWs.** (**A** to **F**) SEM image sequence of Te NWs' growth inside environmental SEM. Most of the growing NWs are planar except for a few, marked in a green circle in image (**F**). At the end of the growth, most of the NWs coalesced to form a layer of Te. The image sequence corresponds to movie S3.

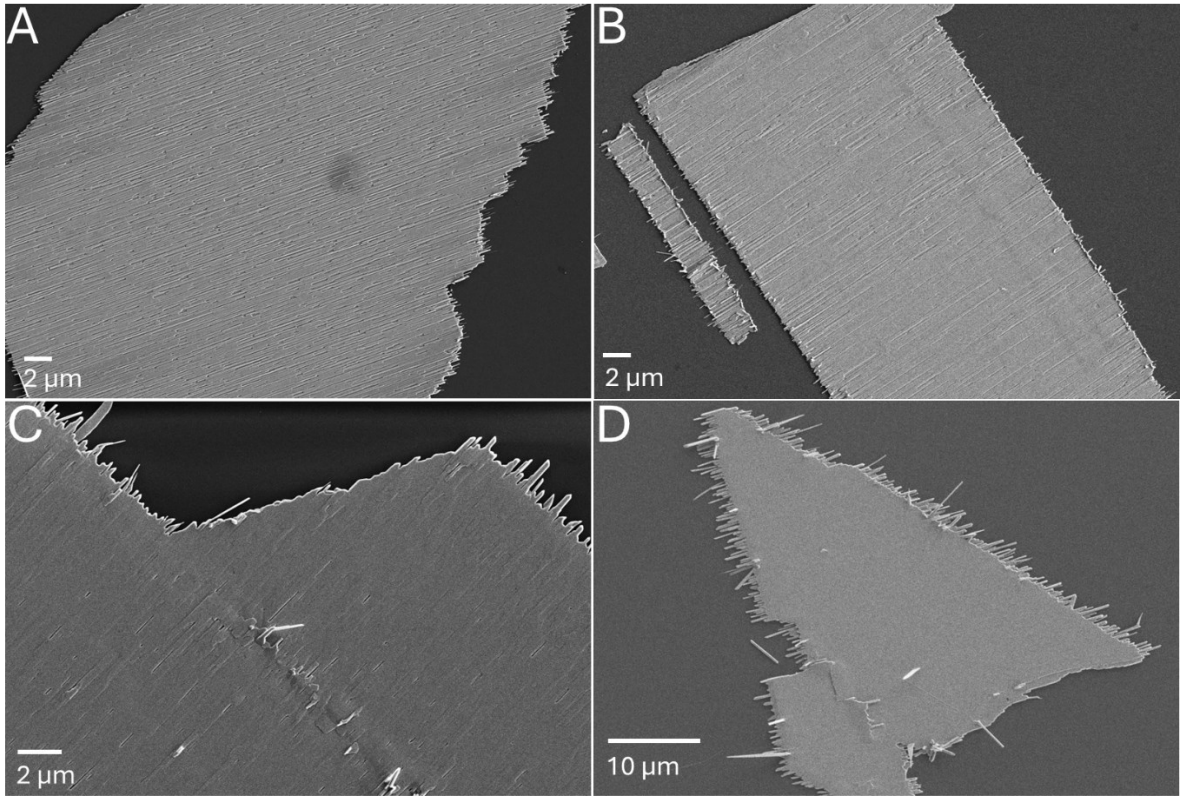

**Fig. S20. SEM image series of ReSe₂ flakes showing the progression of Te coverage**, from densely packed nanowires to a continuous Te film.

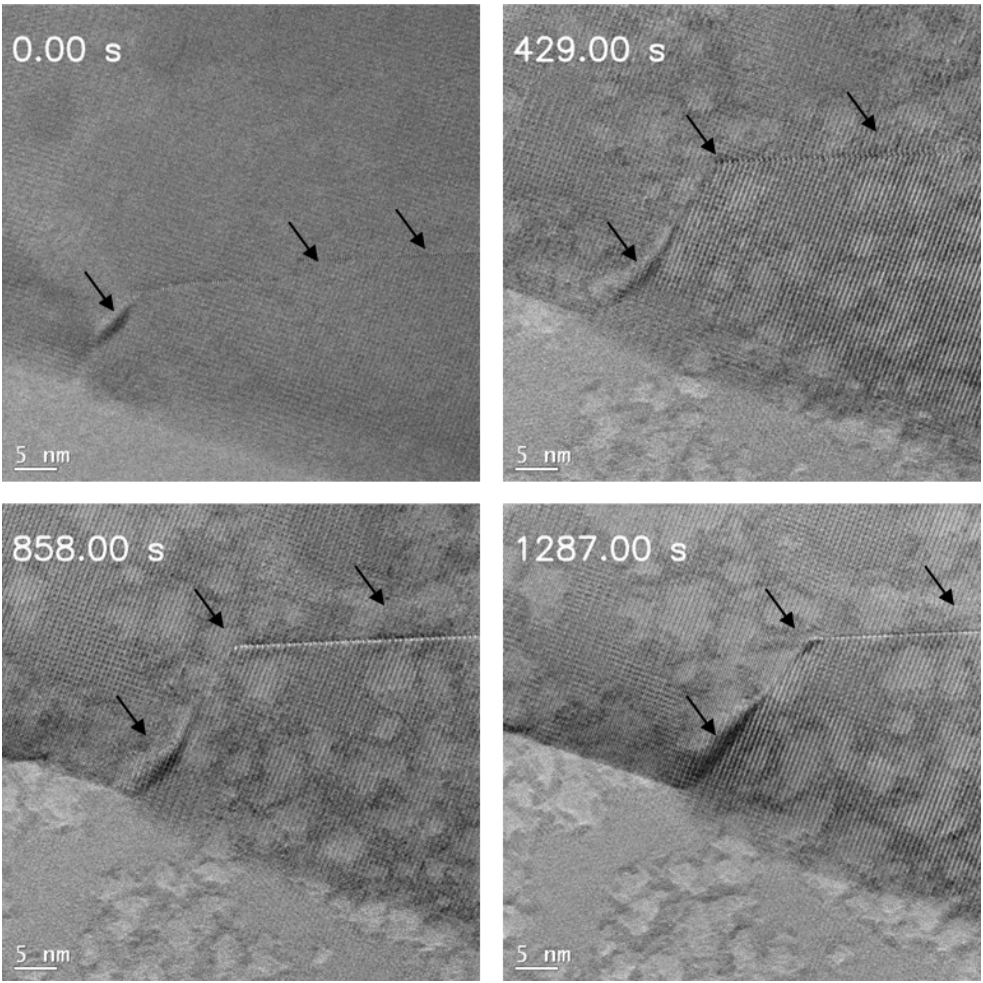

**Fig. S21. Real-time observations of the grain boundary formed between two NWs that grew one towards the other.** TEM image sequence of a grain boundary between two Te NWs growing inside environmental TEM. The image sequence corresponds to movie S7.

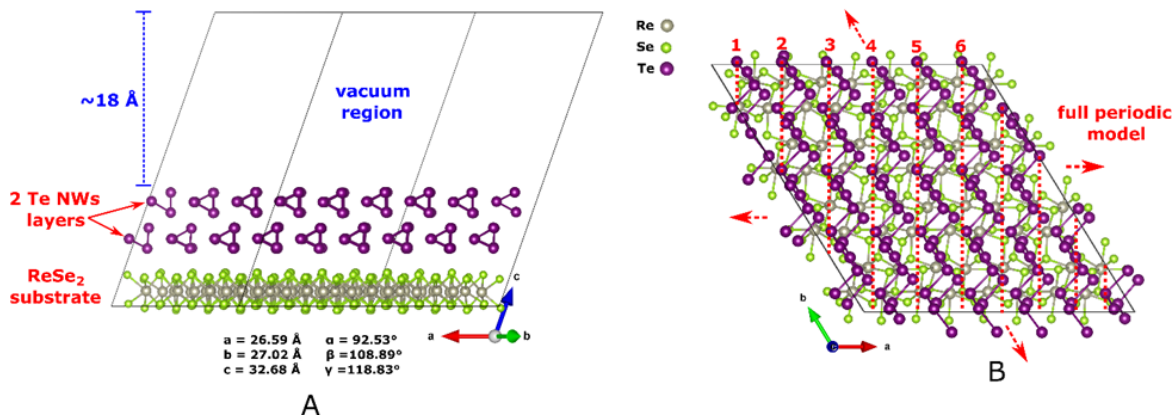

**Fig. S22. Structural model used for *ab initio* DFT calculations**. (A) Side view of the slab geometry for Te chains on the ReSe$_2$ (001) and (00$\bar{1}$) surfaces, including two layers of Te chains and a vacuum region of approximately 18 Å along the z direction to avoid interactions between periodic images. The lattice vectors and supercell parameters are indicated. (B) Top view of the fully periodic model on the surface plane, built from a (4 × 4) ReSe$_2$ supercell that accommodates six Te chains per cell (labeled 1 to 6), with periodic boundary conditions along a and b. Atom colors denote Re, Se, and Te as shown in the legend.

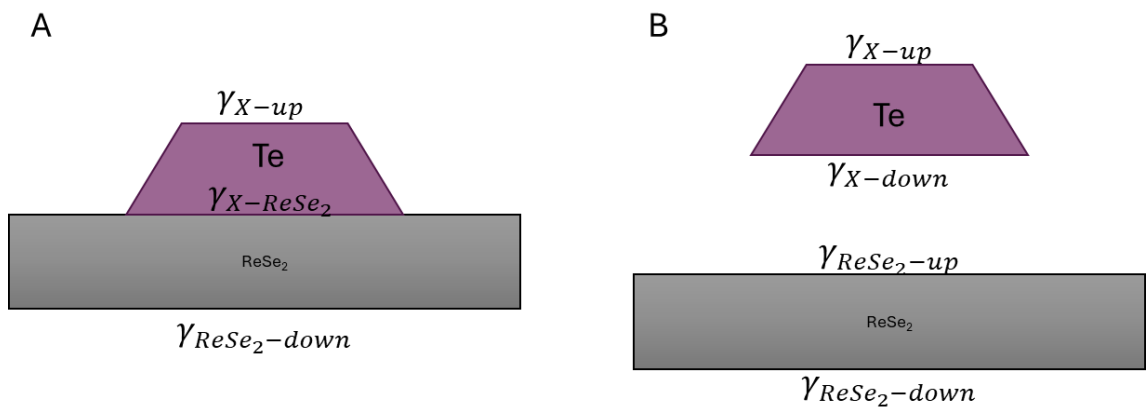

**Fig. S23. Schematic of the surface energies of Te and ReSe₂.** (**A**) Schematic of the surface energies of the Te-ReSe$_2$ heterostructures, the energy at the interface is $\boldsymbol{\gamma_{X-ReSe_2}}$, when X represents a right- or left-handed nucleus. (**B**) Schematic of the surface energy of the same Te and ReSe$_2$ crystals when they are not in contact.

**Table S1. Epitaxial relations of the Te/ReSe$_2$ interfaces**

| L-Te (01$\bar{1}$0) \|\| ReSe$_2$ (001) | L-Te [0001] \|\| ReSe$_2$ [$\bar{1}$20] | L-Te [01$\bar{1}$0] \|\| ReSe$_2$ [C*] |
|---|---|---|
| L-Te (01$\bar{1}$0) \|\| ReSe$_2$ (001) | L-Te [000$\bar{1}$] \|\| ReSe$_2$ [$\bar{1}$20] | L-Te [01$\bar{1}$0] \|\| ReSe$_2$ [C*] |
| L-Te (01$\bar{1}$0) \|\| ReSe$_2$ (00$\bar{1}$) | L-Te [0001] \|\| ReSe$_2$ [120] | L-Te [01$\bar{1}$0] \|\| ReSe$_2$ [-C*] |
| L-Te (01$\bar{1}$0) \|\| ReSe$_2$ (00$\bar{1}$) | L-Te [000$\bar{1}$] \|\| ReSe$_2$ [120] | L-Te [01$\bar{1}$0] \|\| ReSe$_2$ [-C*] |
| R-Te (01$\bar{1}$0) \|\| ReSe$_2$ (001) | R-Te [0001] \|\| ReSe$_2$ [$\bar{1}$20] | R-Te [01$\bar{1}$0]\|\| ReSe$_2$ [C*] |
| R-Te (01$\bar{1}$0) \|\| ReSe$_2$ (001) | R-Te [000$\bar{1}$] \|\| ReSe$_2$ [$\bar{1}$20] | R-Te [01$\bar{1}$0] \|\| ReSe$_2$ [C*] |
| R-Te (01$\bar{1}$0) \|\| ReSe$_2$ (00$\bar{1}$) | R-Te [0001] \|\| ReSe$_2$ [120] | R-Te [01$\bar{1}$0] \|\| ReSe$_2$ [-C*] |
| R-Te (01$\bar{1}$0) \|\| ReSe$_2$ (00$\bar{1}$) | R-Te [000$\bar{1}$] \|\| ReSe$_2$ [120] | R-Te [01$\bar{1}$0] \|\| ReSe$_2$ [-C*] |

Note: C* ≈ [16 9 44]

The table shows the 8 different combinations of Te handedness and orientation, and ReSe$_2$ enantiomorphic basal planes. The C* direction is perpendicular to the basal planes. Due to the triclinic unit cell, this and the opposite directions have large approximate indices (indicated in the note) but are more accurately designated as [C*] or [-C*], respectively.

**Table S2. Frequency of Te nanowires orientation and the handedness on ReSe$_2$**

| Sample # | ReSe$_2$ plane | L-Te [0001] ‖ ReSe$_2$ [$\bar{1}\bar{2}0$] | L-Te [000$\bar{1}$] ‖ ReSe$_2$ [$\bar{1}\bar{2}0$] | R-Te [0001] ‖ ReSe$_2$ [$\bar{1}\bar{2}0$] | R-Te [000$\bar{1}$] ‖ ReSe$_2$ [$\bar{1}\bar{2}0$] |
|---|---|---|---|---|---|
| 1 | (001) | 9 | 10 | 1 | 2 |
| 2 | (001) | 27 | 10 | 1 | 8 |
| 6 | (00$\bar{1}$) | 9 | 9 | 23 | 26 |

Note that in sample #2, Te NWs with orientation Te [0001] ‖ ReSe$_2$ [$\bar{1}\bar{2}0$] have a particularly high enantioselectivity of 27 left-handed Te NWs vs 1 right-handed Te NW, corresponding to 96% L-Te vs. 4% R-Te, and an enantiomeric excess of *EE* = 93%.

Table S3. The calculated surface energy differences between left- and right-handed nuclei

| ReSe$_2$ basal plane | Te handedness | First energy minimum point [eV] | Second energy minimum point [eV] | $\Delta\gamma_{LR}$ [eV] | $\Delta\gamma_{LR}$ [eV Å$^{-2}$] |
|---|---|---|---|---|---|
| (001) | Left | -71161.435 | -71159.053 | -0.096 | $-1.525\cdot10^{-4}$ |
|  | Right | -71158.871 | -71161.339 |  |  |
| (00$\bar{1}$) | Left | -71159.008 | -71161.335 | 0.094 | $1.493\cdot10^{-4}$ |
|  | Right | -71161.429 | -71158.995 |  |  |

**Captions for Movies S1 to S8**

**Movie S1.**

Real-time growth of Te micro- and nanotubes on a ReSe$_2$ flake, when using fast heating rates of 25 °C per minute, recorded by *in situ* SEM. These thick crystals show less selective alignment than thinner NWs (see fig. S17).

**Movie S2.**

Real-time formation of a Te microtube growing on the silicon substrate, grown when using fast heating rates of 25 °C per minute, recorded by *in situ* SEM. These very thick crystals show less selective deposition, so they can grow on the silicon substrate instead of on the ReSe$_2$ flakes. Some of them grow vertically or at an angle out of the substrate plane (see fig. S18).

**Movie S3.**

Real-time growth of Te NWs on ReSe$_2$ recorded by *in situ* SEM showing how the Te NWs merge and coalesce to form a Te film and non-planar growth (see fig. S19). The NWs growth is achieved when using low heating rates of 10 °C per minute.

**Movie S4.**

Real-time growth of aligned Te NWs on ReSe$_2$ recorded by *in situ* SEM at low heating rates of 10 °C per minute. NWs are seen to nucleate first on surface defects like step edges (top left) and later on the basal plane (center).

**Movie S5.**

The same movie as S4, with the overlaid tracking of the NWs, as seen in the image sequence in Fig. 3A. The NWs are shown to elongate simultaneously at their two ends.

**Movie S6.**

Low-magnification real-time growth of several Te NWs recorded by *in situ* TEM, showing nucleation, growth and coalescence of Te NWs. Nucleation is visible from t=6s to t~500s (with the frames up to t~170s appearing noisier due to lower beam intensity), followed by growth. Coalescence events are visible from t~3000s. Diffraction contrast varies across the imaged area due to small changes in sample tilt, and NWs are most visible near the bottom of the frame. The movie shows that coalescence events are the origin of the boundaries visible within individual NWs at the end of the growth experiment. The NWs formed at substrate temperature of 180 °C, recorded in bright field conditions at 4 frames per second and speeded up by 85x.

**Movie S7.**

Growth of a Te NW at substrate temperature 185 °C, recorded in bright field conditions at 1 frame per second and speeded up 73x. This movie yielded the still images in Fig. 3D. The fringes along the NW correspond to the Te covalent chain direction and the contrast gradient near the NW edges indicates its trapezoidal cross-section. Growth is by step flow on the sides, front and top of the NW, and 2-3 steps flow simultaneously as the NW elongates. A change in tip faceting occurs after growth crosses between two regions of the ReSe$_2$ flake showing different contrast.

**Movie S8.**

Real-time observation of the grain boundary dynamics between two NWs that grew towards each other and coalesced, similar to what is shown in Movie S6 near the bottom of the frame. The movie was recorded by *in situ* in the TEM at 190 ℃ in bright field conditions at 1 frame per second and speeded up 86x. Note how the grain boundary fluctuates without disappearing (See fig. S21).